\documentclass{article}
\usepackage{graphicx}
\usepackage{color}
\usepackage{siunitx} 
\usepackage{authblk} 
\usepackage{soul}
\newcommand{\expnumber}[2]{{#1} \! \cdot \! 10^{#2}} 

\begin{document}

\newcommand{\doubleexpectation}{E^{URLT}_{URLA}[C]}
\newcommand{\doubleexpectationwithoutstars}{E^{URLT}_{URLA}[C | \neg \mbox{star}]}

\newcommand{\changed}[1]{{#1}}
\newcommand{\alsochanged}[1]{{#1}}

\title{Are crossing dependencies really scarce?}
\author[1,$*$]{R. Ferrer-i-Cancho}
\author[2]{C. G\'omez-Rodr\'iguez}
\author[3]{J. L. Esteban}
\affil[1]{\small Complexity \& Quantitative Linguistics Lab, LARCA Research Group\authorcr
Departament de Ci\`encies de la Computaci\'o\authorcr
Universitat Polit\`ecnica de Catalunya\authorcr
Campus Nord, Edifici Omega, Jordi Girona Salgado 1-3\authorcr
08034 Barcelona, Catalonia (Spain)}
\affil[2]{Universidade da Coru\~na\authorcr
FASTPARSE Lab, LyS Research Group\authorcr
Departamento de Computaci\'on\authorcr 
Facultade de Inform\'atica, Elvi\~na\authorcr 
15071 A Coru\~na, Spain}
\affil[3]{Logic and Programming, LOGPROG Research Group\authorcr
Departament de Ci\`encies de la Computaci\'o\authorcr
Universitat Polit\`ecnica de Catalunya\authorcr
Campus Nord, Edifici Omega, Jordi Girona Salgado 1-3\authorcr
08034 Barcelona, Catalonia (Spain)}
\affil[*]{Corresponding author, rferrericancho@cs.upc.edu. 
}

\maketitle

\begin{abstract}
The syntactic structure of a sentence can be modelled as a tree, where vertices correspond to words and edges indicate syntactic dependencies.
It has been claimed recurrently that the number of edge crossings in real sentences is small. However, a baseline or null hypothesis has been lacking. 
Here we quantify the amount of crossings of real sentences and compare it to the predictions of a series of baselines. We conclude that crossings are really scarce in real sentences. Their scarcity is unexpected by the hubiness of the trees. Indeed, real sentences are close to linear trees, where the potential number of crossings is maximized. 
\end{abstract}


\section{Introduction}

\label{introduction_section}

Central to network theory is the definition of null models that shed light on the nature and the significance of network properties \cite{Cohen2010a, Newman2010a}. A prototypical example of measure is ${\cal T}$, the clustering coefficient of a network (the average proportion of pairs of neighbours of a vertex that are connected) \cite{Newman2003b}. 
It is well known that real networks typically exhibit ${\cal T} \gg {\cal T}_{ER}$, where ${\cal T}_{ER}$ is the clustering coefficient of an Erd\H{o}s-R\'enyi graph with the same density of links \cite{Newman2003b}. In this setup, ${\cal T}_{ER} = \delta$ where $\delta$ is the density of links of the real network. As real networks are sparse, $\delta$ is a small number while ${\cal T}$ is typically a large number (a number close to $1$). Hence ${\cal T}$ is much greater than expected by chance. 

As a null hypothesis, the Erd\H{o}s-R\'enyi graph involves minimal information from a real network: its number of vertices and its number of links. In an attempt to understand the origins of the properties of real networks, researchers have been defining null hypotheses that are stronger than the Erd\H{o}s-R\'enyi graph in the sense that they involve more information from a real network. Perhaps the most popular example are random graphs with a degree sequence that matches that of the real graph. Various models that differ in how they sample the space of possible graphs have been designed. One is the configuration model or pairing model, a model that has been very successful from a theoretical perspective \cite{Bender1978a, Molloy1995a,Newman2001d} but has very limited applicability as a baseline for real networks due to its inefficiency \cite{Newman2010a}. The configuration model samples uniformly on the space of configurations (pairings of stubs) \cite{Newman2010a}. Another example is the switching model, a model that produces a random graph from a given graph preserving the degree sequence as in the configuration model \cite{Milo2002a,Milo2003a,Maslov2002a}. 
The switching model can be configured to sample uniformly over the space of possible graphs with the same degree sequence \cite{Coolen2009a,Roberts2012a}.

\begin{figure}
\begin{center}
\includegraphics[scale = 0.9]{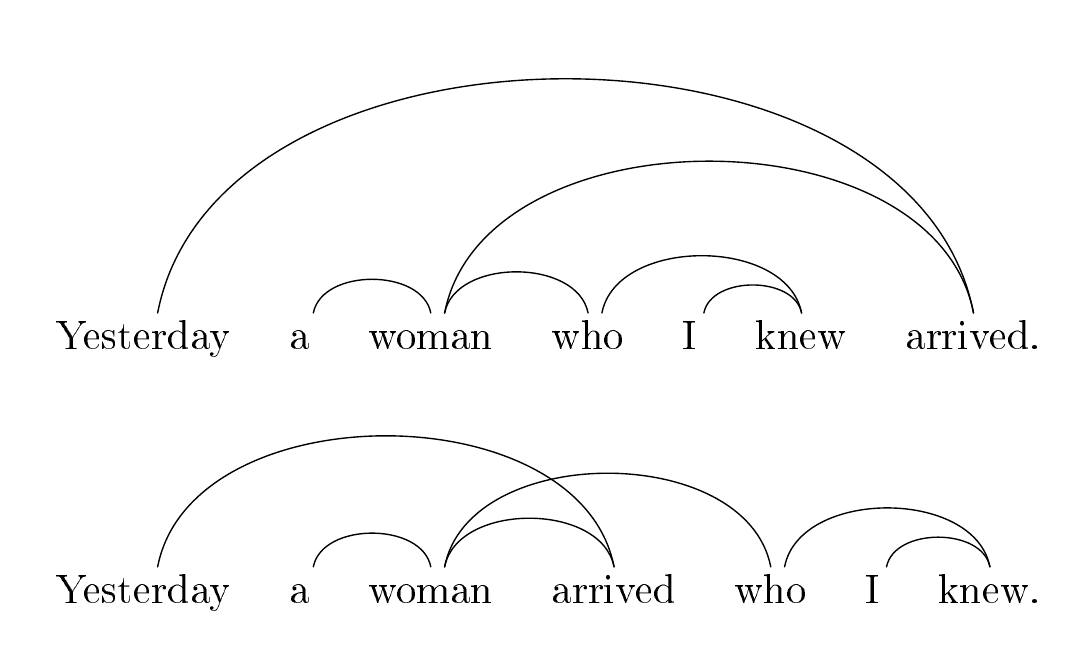}
\end{center}
\caption{\label{extraposition_figure} Top: The syntactic dependency structure of a sentence without crossings. Bottom:
The syntactic dependency structure of the same sentence with an ordering that produces one crossing. Borrowed from \cite{Levy2012a}. 
}
\end{figure}

Here we focus on baselines and null hypotheses for a particular kind of network, i.e., the syntactic structure of sentences, where nodes correspond to words and connections indicate syntactic dependencies between elements, e.g., the dependency between the subject of a sentence and the corresponding verb (Fig. \ref{extraposition_figure}) \cite{melcuk88}. \changed{Syntactic dependency networks are typically trees \cite{tesniere59,hays64,melcuk88} and constitute a particular case of spatial or geographical network \cite{Gastner2006a,Barthelemy2006a,Reuven2010a_Chapter8} in one dimension, the dimension defined by the linear order of the words in the sentence}. 
The specific measure which we aim to compare against null hypotheses is the number of crossings, which we will denote by $C$ in general. Suppose that vertices are arranged sequentially and that $\pi(v)$ is the position of vertex $v$ in the sequence ($\pi(v) = 1$ for the first vertex of the sequence, $\pi(v) = 2$ for the second vertex of the sequence, and so on). Suppose that we have two edges $u \sim v$ and $s\sim t$ such that $\pi(u) < \pi(v)$ and $\pi(s) < \pi(t)$. We say that $u \sim v$ and $s\sim t$ cross if and only if 
$\pi(u) < \pi(s) < \pi(v) < \pi(t)$ or $\pi(s) < \pi(u) < \pi(t) < \pi(v)$.
With this definition one can count $C_{true}$, the observed number of crossings for a given sentence.  
The top of Fig. \ref{extraposition_figure} shows a planar sentence, i.e., a sentence without crossings ($C_{true} = 0$), whereas the bottom shows an ordering of the same sentence with one dependency crossing ($C_{true} = 1$), involving the dependency between ``Yesterday'' and ``arrived'' and the dependency between ``woman'' and ``who''.

\begin{figure}
\begin{center}
\includegraphics{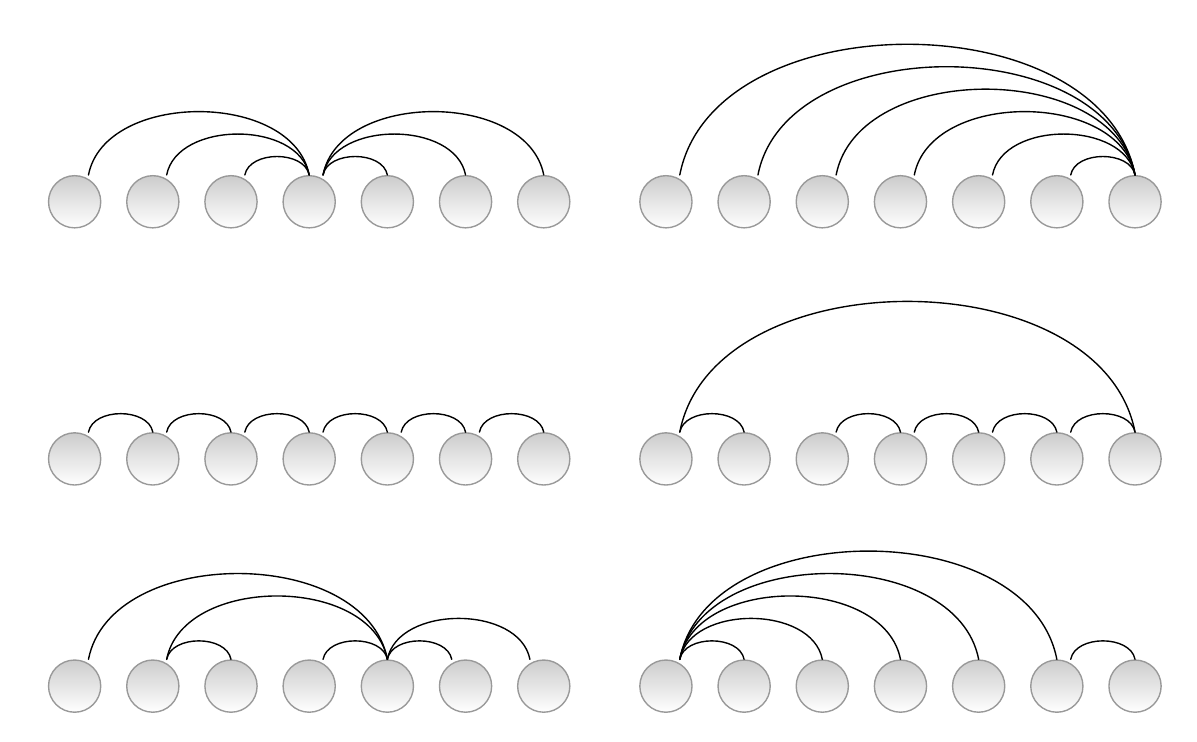}
\caption{\label{star_and_linear_trees_figure} Linear arrangements of trees with $n = 7$ vertices. Top: A star tree. Center: A linear tree. Bottom: A quasi-star tree. }
\end{center}
\end{figure}

It is well known that crossing dependencies, those that cross each other when drawn above the words of a sentence, are relatively uncommon in natural language \cite{lecerf60,hays64}.  
It is widely accepted that the number of crossings of real sentences is small \cite{lecerf60,hays64,melcuk88,Ferrer2006d,Park2009a,Liu2010a,Gildea2010a}. A challenge for the belief that the number of crossings is really small is that the proportion of sentences of a corpus that are not planar, namely, they have at least one crossing, can be very large. For instance, about $30\%$ of sentences in German and Dutch corpora are not planar (Table 1 of \cite{GomNiv2013}). Another challenge is that the scarcity of crossings is not supported with a baseline or null hypothesis. For instance, a star tree (a tree where all connections are formed with a hub vertex as on top of Fig. \ref{star_and_linear_trees_figure}) cannot have crossings \cite{Ferrer2013b}. Therefore, reaching the theoretical minimum number of crossings does not suffice to conclude that the number of crossings is smaller than expected by chance: for a star tree, it could not be otherwise. In other words, the number of crossings of a star tree is really small (it is minimum) but not scarce with respect to all possible linear orderings of its vertices.  
 
In this article, we aim to quantify the actual number of crossings of real sentences and to clarify the issue of the presumable scarcity of crossing dependencies. 
More specifically, we will calculate the actual number of crossings in large collections of sentences and compare them against the predictions of baselines and null hypotheses that vary in the amount of information that they involve about a real tree, as it happens with null hypotheses for real networks. 

The remainder of the article is organized as follows. Section \ref{baselines_section} presents a series of baselines that will be used to assess if the actual number of crossings in sentences is really scarce. Some baselines are borrowed from previous research \cite{Ferrer2014c,Ferrer2013d} while others are introduced here. It also presents a measure of hubiness (a normalized measure of the similarity between a dependency tree and a star tree) and shows the relationship between that measure and the potential number of crossings of a tree.
Section \ref{materials_section} presents the collections of dependency networks from different languages that will be used in Section \ref{results_section} to compare the actual number of crossings of real dependency trees against the random baselines of Section \ref{baselines_section}. Section \ref{results_section} also analyzes the degree of hubiness of real dependency trees. 
Section \ref{discussion_section} discusses the results.

\section{Baselines for the number of crossings}

\label{baselines_section}

\subsection{Absolute baselines for the number of crossings}

\changed{Star trees and linear trees are crucial to understand the limits of the variation of $C$. }
A star tree is a tree where a vertex has maximum degree (namely $n - 1$, an thus all other vertices have degree 1) \cite{Ferrer2013b}. A linear tree is a tree where vertex degrees do not exceed 2 (and therefore all vertices have degree 2 except a couple that have degree 1) \cite{Ferrer2013b}. 
See examples of star and linear trees in Fig. \ref{star_and_linear_trees_figure}.

When looking for a reference for the actual number of crossings of a sentence, a first step is to calculate the potential number of crossings. In a syntactic dependency tree of $n$ nodes, the number of edges is 
$n-1$ and therefore the total number of crossings cannot exceed
\begin{equation}
{n - 1 \choose 2} = \frac{(n - 1)(n - 2)}{2}.
\label{wild_potential_number_of_crossings_of_equation}
\end{equation}
However, this is a rough estimate as edges that share a vertex cannot cross. Taking account this fact, one may define $Q$, the size of the set of pair of edges that may potentially cross.
$|Q|$, the cardinality of this set, depends on $n$ and $\left< k^2 \right>$, the second moment of degree about zero, defined as 
\begin{equation}
\left< k^2 \right> = \frac{1}{n} \sum_{i=1}^n k_i^2,
\end{equation}
with $k_i$ being the degree of the $i$-th vertex of the network. In particular, one has \cite{Ferrer2013b} 
\begin{equation}
|Q| = \frac{n}{2}\left(\left< k^2 \right>_{star} - \left< k^2 \right>\right),
\label{potential_number_of_crossings_equation}
\end{equation} 
where 
\begin{equation}
\left< k^2 \right>_{star} = n - 1 
\label{degree_2nd_moment_star_tree_equation}
\end{equation}
is the value of $\left< k^2 \right>$ for a star tree of $n$ vertices \cite{Ferrer2013b}. Indeed, 
$|Q|$ reaches extreme values in star trees and linear trees.
An overview of the arguments follows (see \cite{Ferrer2013d} and the Appendix of  \cite{Ferrer2015c} for further mathematical details).

The variation of $|Q|$ obeys 
\begin{equation} 
|Q_{star}| \leq |Q| \leq |Q_{linear}|,
\end{equation}
where $|Q_{linear}|$ and $|Q_{star}|$ are the value of $|Q|$ in a linear tree and a star tree, respectively. Obviously, the minimum $|Q|$, namely, $|Q| = 0$ is achieved by a star tree because 
$\left< k^2 \right> = \left< k^2 \right>_{star}$ in that case. The maximum value of $|Q|$ is achieved by a linear tree because that tree yields the minimum value of $\left< k^2 \right>$, namely 
\begin{equation}
\left< k^2 \right>_{linear} = 4 - 6/n
\label{degree_2nd_moment_linear_tree_equation}
\end{equation} 
(when $n \geq 2$). Notice that  
\begin{equation}
\left< k^2 \right>_{star} - \left< k^2 \right>_{linear} = \frac{(n-2)(n-3)}{n}
\label{maximum_difference_degree_2nd_moment_equation}
\end{equation}
Therefore, the maximum value of $|Q|$ is 
\begin{eqnarray}
|Q_{linear}| & = & \frac{n\left(\left< k^2 \right>_{star} - \left< k^2 \right>_{linear}\right)}{2} \nonumber \\
             & = & \frac{1}{2}(n-2)(n-3) 
\label{potential_number_of_crossings_of_linear_tree_equation}
\end{eqnarray}
for $n \geq 3$ ($|Q_{linear}| = 0$ for $n < 3$). 

\begin{figure}
\includegraphics{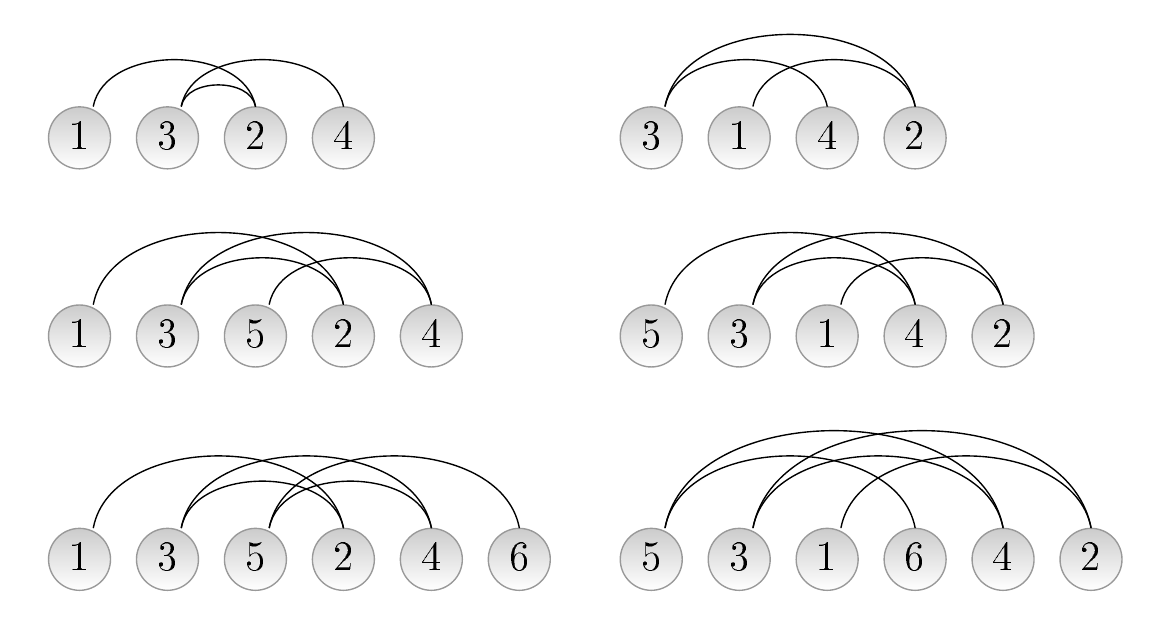} 
\caption{\label{extremal_arrangements_small_linear_trees_figure} Arrangements of small linear trees with maximum $C$ according to Eq. \ref{potential_number_of_crossings_of_linear_tree_equation}. Top: $n = 4$ and $C = |Q_{linear}| = 1$.
Center: $n = 5$ and $C = |Q_{linear}| = 3$. Bottom: $n = 6$ and $C = |Q_{linear}| = 6$.   
}
\end{figure}

\changed{Obviously, $C = |Q_{star}| = 0$ for any linear arrangement of the vertices of a star tree. We wish to check if there are orderings of the vertices of a linear tree where actually $C = |Q_{linear}|$. }
Suppose that the vertices of a linear tree with $n \geq 2$ are labelled following a depth-first traversal from one of the leaves. This is equivalent to labelling vertices according to their position in a minimum linear arrangement \cite{Esteban2016a}. Fig. \ref{extremal_arrangements_small_linear_trees_figure} shows linear arrangements of small linear trees with maximum $C$, namely $C = |Q_{linear}|$. These arrangements can be built by placing all vertices with odd labels in ascending order followed by all vertices with even labels in ascending order. By symmetry, it is possible to build other arrangements where $C= |Q_{linear}|$, e.g., placing all vertices with odd labels in descending order followed by all vertices with even labels in descending order. Appendix \ref{maximum_crossings_linear_tree_appendix} presents arrangements that reach $C = |Q_{linear}|$ for a linear tree of an arbitrary size $n$ ($n \geq 3$). 
As 
\begin{equation} 
|Q_{linear}| = {n - 2 \choose 2},
\end{equation}
(recall Eq. \ref{potential_number_of_crossings_of_linear_tree_equation}), 
it is easy to see how crossing theory replaces the naive upper bound of $C$ in Eq. \ref{wild_potential_number_of_crossings_of_equation} with a tight one. 

Given the results above, the actual number of crossings of a star tree is not surprising at all according to $|Q|$. In contrast, achieving a low number of crossings in a linear tree is unexpected if the tree is sufficiently large. 

\subsection{A hubiness coefficient}

We have seen above that $\left<k^2 \right>$ is a fundamental structural property of a tree: it determines $|Q|$. As $\left<k^2 \right>$ determines the range of variation of $|Q|$, $\left<k^2 \right>$ also determines the solution to the minimum linear arrangement problem: the solution is minimum for linear trees and maximum for star trees \cite{Esteban2016a}.   
$\left<k^2 \right>$ is a measure of the hubiness of a tree \cite{Ferrer2013b} and we have seen that its range of variation is 
\begin{equation} 
\left<k^2 \right>_{linear} \leq \left<k^2 \right> \leq \left<k^2 \right>_{star},
\label{k2_variability_range_equation}
\end{equation}
where $\left<k^2 \right>_{linear}$ and $\left<k^2 \right>_{star}$ are the value of $\left<k^2 \right>$ in a linear tree and a star tree respectively.
The latter allows one to define a hubiness coefficient $h$ as
\begin{equation} 
h = \frac{\left<k^2 \right> - \left<k^2 \right>_{linear}}{\left<k^2 \right>_{star}-\left<k^2 \right>_{linear}}
\label{hubiness_coefficient_equation}
\end{equation}
for $n \geq 4$ (for $n < 4$, the only trees that can be formed are both linear and star trees).
It is easy to show that $0 \leq h \leq 1$. On the one hand, the fact that  
\begin{eqnarray}
\left<k^2 \right> - \left<k^2 \right>_{linear} & \geq & \left<k^2 \right>_{linear} - \left<k^2 \right>_{linear} \nonumber \\
                                               & = & 0
\end{eqnarray}
gives
\begin{equation}
h_{linear} = 0 \leq h.
\end{equation}
On the other hand, the fact that  
\begin{equation}
\left<k^2 \right> - \left<k^2 \right>_{linear} \leq \left<k^2 \right>_{star} - \left<k^2 \right>_{linear}
\end{equation}
gives
\begin{equation}
h \leq h_{star} = 1.
\end{equation}
Therefore, $h$ measures the similarity between a tree and a star tree (or the dissimilarity with respect to a linear tree) from the perspective of $\left<k^2 \right>$.
Applying Eqs. \ref{degree_2nd_moment_star_tree_equation} and \ref{degree_2nd_moment_linear_tree_equation} to Eq. \ref{hubiness_coefficient_equation}, one obtains 
\begin{eqnarray} 
h & = & \frac{n\left(\left<k^2 \right> - 4\right) + 6}{n^2-5n+6} \nonumber \\
  & = & \frac{n\left(\left<k^2 \right> - 4\right) + 6}{(n - 2)(n - 3)}. \label{final_hubiness_coefficient_equation}
\end{eqnarray} 

\changed{Note that $h$ is a normalized degree variance. To see it, recall that the degree variance is
\begin{equation}
V[k] = \left< k^2 \right> - \left<k\right>^2
\end{equation}
and that $V[k]$ is fully determined by $\left< k^2 \right>$ because $\left<k\right>= 2 - 2/n$ \cite{Noy1998a} for any tree such that $n \geq 1$. Therefore, 
\begin{equation}
V[k]_{linear} \leq V[k] \leq V[k]_{star}
\end{equation}
and 
\begin{equation}
h = \frac{V[k] - V[k]_{linear}}{V[k]_{star} - V[k]_{linear}}.
\end{equation}
$h$ is 1 when the degree variance is maximum and 0 when variance is minimum. 
}

It is \changed{also} easy to show that $h$ is the complementary of the normalized potential number of crossings, i.e.,  
\begin{equation}
h = 1- \frac{|Q|}{|Q_{linear}|}.
\label{hubiness_vs_potential_crossings_equation}
\end{equation}
\changed{Therefore, $h$ is 1 when the potential number of crossings is minimum and 0 when it is maximum. }
Applying the definition of $|Q|$ in Eq. \ref{potential_number_of_crossings_equation} and $|Q_{linear}|$ in Eq. \ref{potential_number_of_crossings_of_linear_tree_equation} to
\begin{equation}
h = \frac{|Q_{linear}| - |Q|}{|Q_{linear}|},
\end{equation} 
one recovers Eq. \ref{final_hubiness_coefficient_equation} after some algebra.


\subsection{Random baselines for the number of crossings}

\label{random_baselines_section}

We consider $C$, the number of crossings of a sentence, in a uniformly random linear arrangement (URLA) of its  elements. In this baseline, the expected number of crossings is \cite{Ferrer2014c, Ferrer2013d}
\begin{equation}
E_{URLA}[C] = \frac{|Q|}{3}.
\label{expected_number_of_crossings_equation}
\end{equation} 
Another baseline can be obtained assuming that the tree is a uniformly random labelled tree (URLT).
\changed{This choice improves previous research where random trees that deviate from a uniform distribution were used \cite{Ferrer2006d} as a control for $C_{true}$. }

The expected value of $\left<k^2\right>$ in a URLT is \cite{Ferrer2014c}
\begin{eqnarray}
E_{URLT}\left[ \left<k^2\right> \right] & = & \left(1-\frac{1}{n}\right) \left(5-\frac{6}{n} \right) \nonumber \\
                                        & = & \frac{(n-1)(5n - 6)}{n^2}\label{expected_degree_2nd_moment_equation}
\end{eqnarray}
and then the expected number of crossings in a URLA of an URLT is
\begin{eqnarray}
\doubleexpectation & = & E_{URLA}\left[\frac{|Q|}{3}\right] \nonumber \\
                   & = & \frac{n}{6} \left(n - 1 - E\left[\left<k^2\right>\right] \right) \label{expected_crossings_equation} \\
                                   & = & \frac{(n-1)(n-2)(n-3)}{6n}.
\label{expected_number_of_crossings_of_uniformly_random_tree_equation}
\end{eqnarray}
Notice that $\doubleexpectation$ is related with the unrestricted baseline of the previous subsection. Combining Eqs. \ref{potential_number_of_crossings_of_linear_tree_equation} and \ref{expected_number_of_crossings_of_uniformly_random_tree_equation}, one obtains
\begin{eqnarray}
\doubleexpectation & = & \frac{n-1}{3n}|Q_{linear}| \label{raw_expected_number_of_crossings_URLT_equation}\\
                                   & \approx & \frac{1}{3}|Q_{linear}| \label{expected_number_of_crossings_URLT_equation}
\end{eqnarray}
for sufficiently large $n$. This implies that the expected number of crossings in a random linear arrangement of a URLT is very close to the expected number of crossings in a URLA of a linear tree.
 
\subsection{Random baselines for the hubiness coefficient}

Recalling Eqs. \ref{maximum_difference_degree_2nd_moment_equation} and \ref{degree_2nd_moment_linear_tree_equation}, it is easy to see that the expected value of the hubiness coefficient in a uniformly random labelled tree is 
\begin{eqnarray} 
E_{URLT}[h] & = & E\left[\frac{\left<k^2 \right> - \left<k^2 \right>_{linear}}{\left<k^2 \right>_{star}-\left<k^2 \right>_{linear}}\right] \nonumber \\
            & = & \frac{n\left(E_{URLT}\left[\left<k^2 \right>\right] - \left<k^2 \right>_{linear}\right)}{(n-2)(n-3)}. \label{expected_hubiness_equation}
\end{eqnarray}
Recalling \ref{expected_degree_2nd_moment_equation} and noting that 
\begin{equation}
E_{URLT}\left[\left<k^2 \right>\right] - \left<k^2 \right>_{linear} = \frac{(n-2)(n-3)}{n^2},
\end{equation}
we finally obtain  
\begin{equation}
E_{URLT}[h] = \frac{1}{n}. 
\end{equation}
The latter implies that, as $n$ tends to infinity, the expected hubiness of URLTs vanishes while the similarity between URLTs and linear trees is maximized. Linear trees swallow practically all probability mass, in agreement with the finding that the expected number of crossings in a URLA of a URLT tends to that of a linear tree as $n$ tends to infinity (Eq. \ref{expected_number_of_crossings_URLT_equation}). Furthermore, the finding that $E_{URLT}[h] = 1/n$ suggests that the harmonic mean (or its inverse) could be used to evaluate the hubiness of real sentences with respect to URLTs.  

\subsection{Network theory revisited}

In Section \ref{introduction_section}, we have reviewed various null hypotheses that are used in network theory. The Erd\H{o}s-R\'enyi model takes the number of vertices and the number of links of a real network and discards the structure of the real network. The configuration or pairing model and the switching model go a step further incorporating the degree distribution. Our baselines and null hypotheses also parallel this increasing amount of information about the real network that they incorporate. 

Recall the two kinds of upper bounds for $C_{true}$ in Section \ref{baselines_section}. If we consider the structure of the tree under consideration irrelevant (e.g., $\left< k^2 \right>$), the upper bound is $|Q_{linear}|$ (Eq. \ref{potential_number_of_crossings_of_linear_tree_equation}), the maximum value that $|Q|$ can achieve. This bound parallels the Erd\H{o}s-R\'enyi model (notice that in a tree the number of edges is $n-1$ and thus not relevant). 
If we consider the tree structure relevant, then the upper bound is $|Q|$ (Eq. \ref{potential_number_of_crossings_equation}) with $\left< k^2 \right>$ calculated on the tree under consideration. This bound parallels the configuration or pairing model and the switching model: it involves the degree sequence but knowing $\left< k^2 \right>$ suffices. 

Recall also the two kinds of random baselines for $C_{true}$ in Section \ref{random_baselines_section}. Neglecting the structure of the tree under consideration (e.g., $\left< k^2 \right>$), a potential \changed{baseline} is $\doubleexpectation$ (Eq. \ref{expected_number_of_crossings_of_uniformly_random_tree_equation}), the expected number of crossings in a uniformly random tree. This null hypothesis parallels the Erd\H{o}s-R\'enyi model.
Conditioning on the tree structure, then the potential \changed{baseline} is $E_{URLA}[C]$ (Eq. \ref{potential_number_of_crossings_equation}) with $\left< k^2 \right>$ taken from the tree under consideration. This null hypothesis parallels the configuration or pairing model and the switching model for involving the degree sequence or a function of it. 

Our hubiness coefficient is a normalized $\left<k^2 \right>$ and we have seen that $\left<k^2 \right>$ plays a fundamental role in trees: its extremal values determine the limits of the variation of $Q$ and thus also the expected number of crossings in a random linear arrangement of a tree. These values also determine the limits of the variation of $D_{min}$, the minimum sum of edge lengths in a linear arrangement of a given tree ($D_{min}$ is minimum for linear trees and maximum for star trees) \cite{Esteban2016a}. Such a role is reminiscent of the role played by $\left<k^2\right>/\left<k\right>$ in large complex networks concerning, for instance, the spread of epidemics on a network (e.g. a virus on the Internet): if $\left<k^2\right>/\left<k\right>$ diverges the pandemics cannot be stopped \cite{PastorSatorras2004a}. As $\left<k\right> = 2 - 2/n$ in trees \cite{Noy1998a}, our work extends the importance of $\left<k^2\right>/\left<k\right>$ to the domain of trees. 

\section{Materials and methods}

\label{materials_section} 

We aim to compare $C_{true}$ against the different baselines with the help of dependency treebanks. 
A dependency treebank is a collection, or corpus, of sentences where a dependency graph is provided for every sentence. Our treebanks come from version 2.0 of the HamleDT collection of treebanks \cite{HamleDTJournal,HamledTStanford}. This collection harmonizes previously existing treebanks for 30 different languages into two widely-used annotation guidelines: Universal Stanford dependencies \cite{UniversalStanford} and Prague dependencies \cite{PDT20}. Therefore, this resource allows us to evaluate the baselines not only across a wide range of languages of different families, but also across two well-known annotation schemes. This is useful because observations like the number of dependency crossings in a sentence not only depend on the language, as they are also influenced by annotation criteria (\cite{Ferrer2015c} review some examples of how $C$ can be affected by annotation criteria).


As preprocessing, we removed nodes corresponding to punctuation from the analyses in the treebanks, following common practice in research related to statistical properties of dependency structures (e.g. \cite{Ferrer2004b, Futrell2015a}), which is only concerned with dependency relations between actual words. Null elements, which are present in the Bengali, Hindi and Telugu corpora, were also removed as they do not correspond to words. To preserve the structure of the rest of the tree after removing these nodes, non-deleted nodes that had a deleted node as their head were reattached as dependents of their nearest non-removed ancestor. \changed{The size of the tree that is obtained corresponds to the length of the sentence in words. } 


After this preprocessing, we included in our analyses those syntactic dependency structures that (1) defined a tree with at least 4 nodes, and such that (2) the tree was not a star tree. 
The reason for (1) is that our baselines assume a tree structure \cite{Ferrer2013d,Ferrer2014c} and that we wished to avoid the statistical problem of mixing trees with other kinds of graphs, e.g., the potential number of crossings depends on the number of edges \cite{Ferrer2013b,Ferrer2014f,Ferrer2013d}.
We focus on trees of at least 4 nodes because for $n<4$, the number of crossings is always zero.
The reason for (2) is that a star tree cannot have crossing dependencies \cite{Ferrer2013b}. Ratios with $C$ in the numerator and $|Q|$ in the denominator, e.g., the relative number of crossings, $C/|Q|$ \cite{Ferrer2014c}, are not defined because $C = |Q| = 0$.
Tables \ref{hubiness_table_stanford} and \ref{hubiness_table_prague} show $p(star)$, the proportion of trees that are star trees (this proportion is calculated after applying condition (1)). On average, this proportion is smaller than $5\%$.

As star trees are excluded, the random baselines on uniformly random trees must be adapted (see Appendix \ref{expectations_without_star_trees_appendix} for further details). $\doubleexpectation$ is replaced by the same expectation conditioning on the fact that star trees are excluded, i.e. 
\begin{equation}
\doubleexpectationwithoutstars = \frac{(n-1)(n-2)(n-3)}{6(n - n^{4 - n})}. \label{main_adapted_expected_number_of_crossings_of_uniformly_random_tree_equation}
\end{equation}
It is easy to see that 
\begin{equation}
\doubleexpectationwithoutstars \approx \doubleexpectation
\end{equation}
for sufficiently large $n$ (compare Eqs. \ref{expected_number_of_crossings_of_uniformly_random_tree_equation} and \ref{main_adapted_expected_number_of_crossings_of_uniformly_random_tree_equation}).

The same applies to $E_{URLT}[h]$, that has to be replaced by 
\begin{equation}
E_{URLT}\left[h \middle| \neg \mbox{star} \right] = \frac{n^{n - 4} - 1}{n^{n - 3} - 1}.
\end{equation}
It is easy to see that 
\begin{equation}
E_{URLT}\left[h \middle| \neg \mbox{star}\right] \approx E_{URLT}[h] = \frac{1}{n}
\end{equation}
for sufficiently large $n$.  

The corrected versions of the random baselines are expected to matter especially in treebanks with a sufficient concentration of sentences near $n = 4$. 

\begin{figure}
\includegraphics[scale = 0.7]{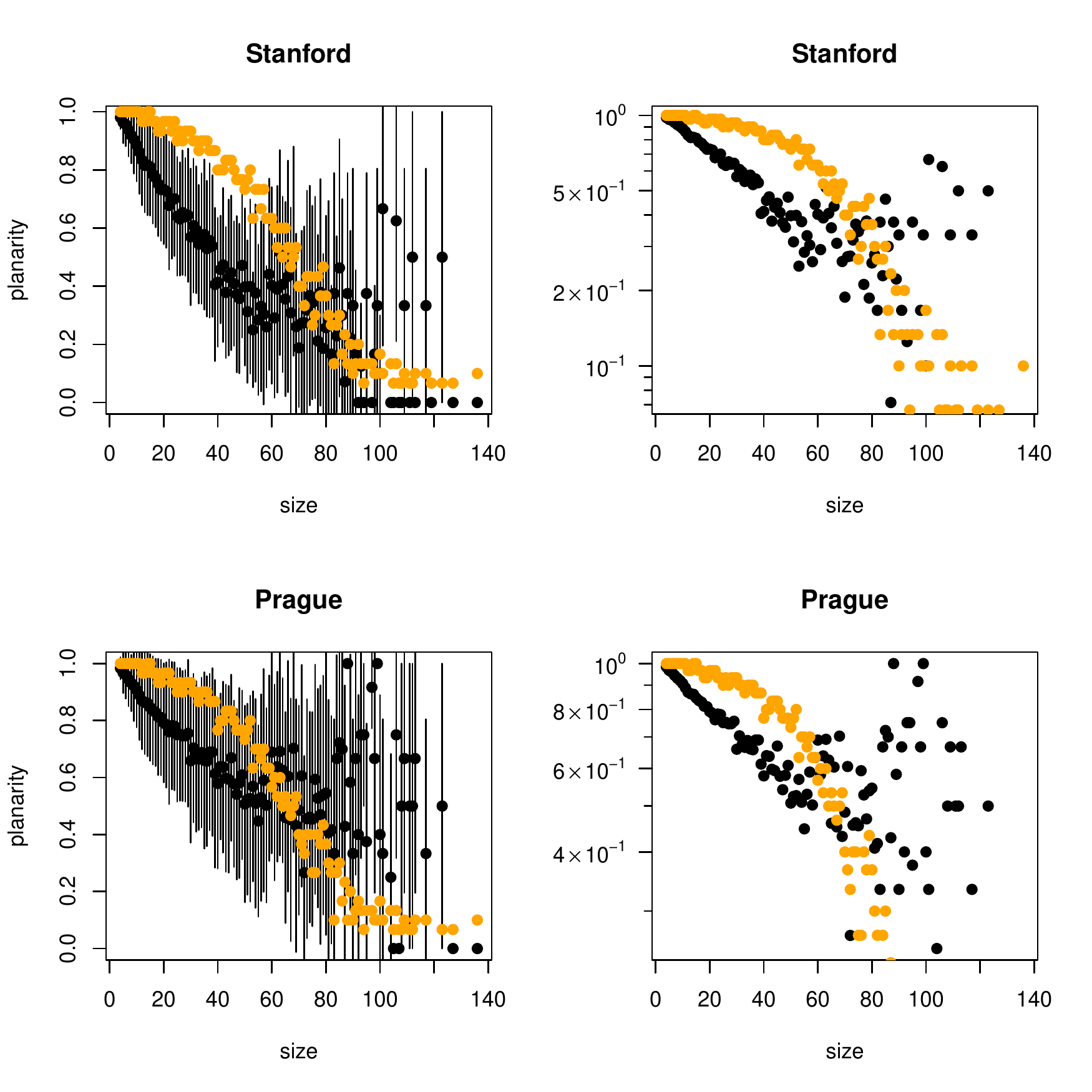}
\caption{\label{planarity_figure} $p(C_{true} = 0)$, the proportion of planar sentences (black), as a function of $n$, the tree size. The proportion of treebanks having at least one tree of size $n$  is also shown (orange). \changed{In the left plots}, points and error bars indicate, respectively, mean values and $\pm 1$ standard deviation over proportions in a collection of treebanks. \changed{In the right plots, error bars are omitted.} Tree sizes represented by less than two treebanks are excluded. Therefore the smallest proportion of treebanks above is $1/15$. Top: Stanford annotations. Bottom: Prague annotations. }
\end{figure}

\changed{To assess if the number of crossings in our dataset is significantly small, we conducted two Monte Carlo tests for each treebank, corresponding to each of the two random models of trees. In the first test, we evaluated the significance of the observed values of $\left< C_{true} \right>$ for each treebank with respect to URLTs, by generating randomized versions of the corpora where each tree is replaced by an URLT with the same number of nodes. To generate each URLT, we produced a uniformly distributed Pr\"ufer code \cite{Pruefer1918a} and then converted it to a tree (as an implementation with the Aldous-Broder algorithm \cite{Aldous1990a,Broder1989a} proved too slow). \alsochanged{The Monte Carlo procedure is used to estimate left-$P$, the probability that a randomized corpus yields a value of $\left< C_{true} \right>$ that is at least as small as the original one. One concludes that $\left< C_{true} \right>$ is significantly small if left-$P$ is small enough.} 
In the second test, we evaluated the significance of $\left< C_{true} \right>$ with respect to URLAs of the trees in the treebank, by generating randomized versions of the treebanks where each syntactic tree is replaced by an URLA of itself. \alsochanged{left-$P$ is estimated as in the 1st test}. Each test is based on $10^4$ randomizations of the treebank. Notice that these tests preserve the distribution of tree sizes of the original treebank, that is required to evaluate the significance of a measurement over a whole treebank accurately \cite{Ferrer2013c}. }  

\changed{We also performed the same \alsochanged{couple of tests} to evaluate the significance of $p(C_{true} = 0)$, the proportion of planar sentences of a treebank. \alsochanged{To evaluate the significance of $\left<h\right>$, we used URLTs as in the 1st test to estimate left-$P$ and also right-$P$ (the latter being the probability that the randomized corpus yields a value of $\left< C_{true} \right>$ that is at least as large as the original one).} } 

\section{Results}

\label{results_section}

The claim that dependency crossings are scarce in real sentences can be evaluated with at least two statistics. Firstly, $p(C_{true} = 0)$, the proportion of planar sentences (sentences without crossings). $p(C_{true} = 0)$ tends to decrease as $n$ increases on average for all treebanks (Fig. \ref{planarity_figure}). A detailed analysis over all tree sizes shows that this number varies substantially across treebanks (Tables \ref{crossings_table_stanford} and \ref{crossings_table_prague}). It is minimum in Ancient Greek with $p(C_{true} = 0) \approx 0.3$ while it reaches its theoretical maximum value ($p(C_{true} = 0) = 1$) for Japanese and Romanian with Prague dependencies. The second smallest proportion of planar sentences is achieved by 
Latin with $p(C_{true} = 0) \approx 0.5$, followed by German and Dutch with $p(C_{true} = 0)$ slightly below $0.7$.  
Our findings are consistent with a previous report of $30\%$ of sentences in German and Dutch that are not planar (Table 1 of \cite{GomNiv2013}). 

\begin{figure}
\includegraphics[scale = 0.7]{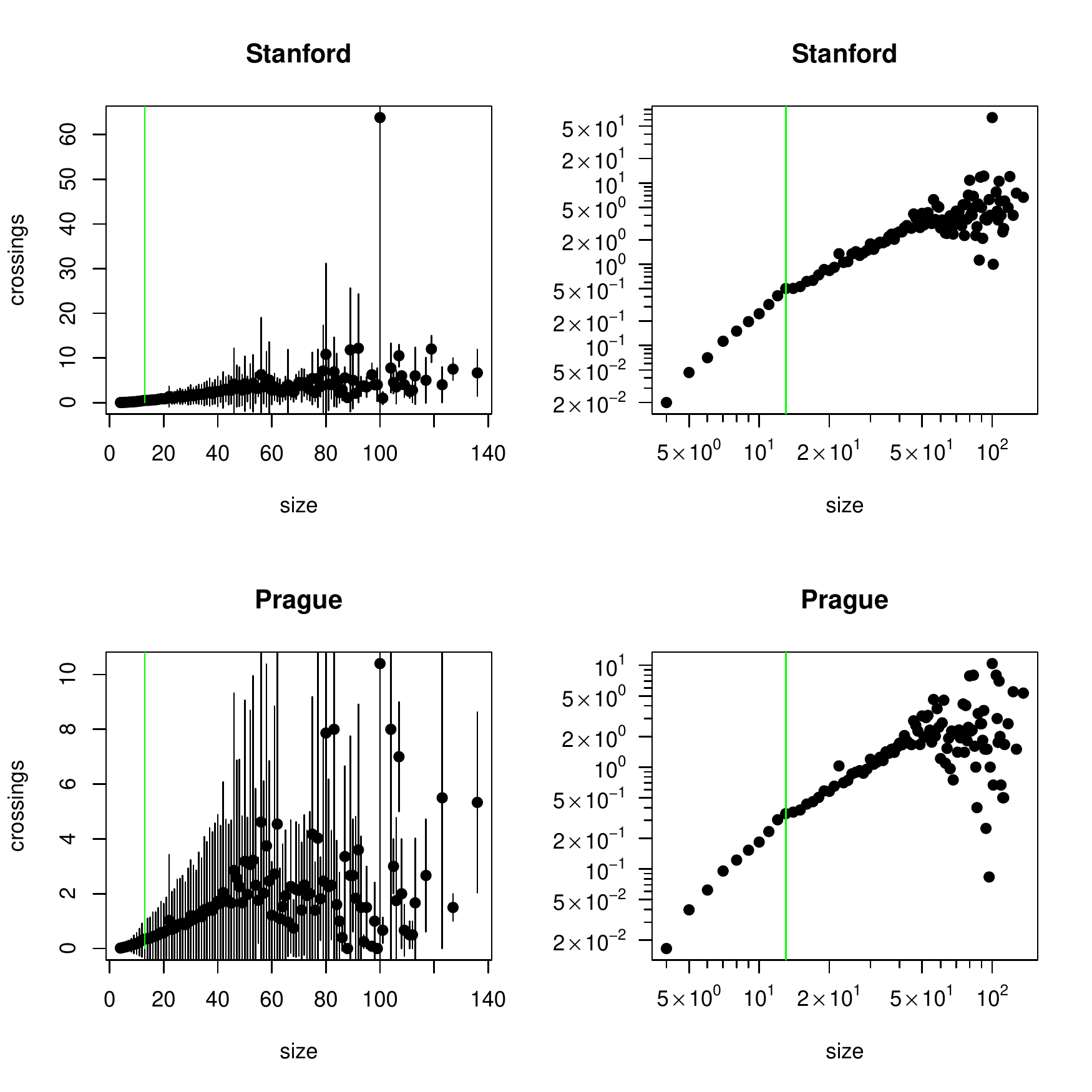}
\caption{\label{crossings_figure} $\left< C_{true} \right>$, the mean number of dependency crossings (black) as a function of $n$, the tree size. As a guide to the eye, a vertical line (green) for $n = 13$ is also shown. Top: Stanford annotations. Bottom: Prague annotations. The format is the same as in Fig. \ref{planarity_figure}.}
\end{figure}

Secondly, one can look at the behavior of the actual number of crossings. $C_{true}$ tends to increase as $n$ increases over all treebanks (Fig \ref{crossings_figure}). Interestingly, the plots in double logarithmic scale reveal the presence of a breakpoint at $n = 13$ that separates an initial regime of fast growth of $C_{true}$ from a second regime of slower growth (Fig \ref{crossings_figure}). 
Hereafter, we will use $\left< ... \right>$ over a tree measure to indicate a mean over the whole ensemble of sentences of a treebank included in our analysis.
Although the proportion of planar sentences can be very low when putting all tree sizes together, the number of crossings is apparently small: $\left<C_{true}\right>$ does not reach $3.4$ in any of the treebanks (Tables \ref{crossings_table_stanford} and \ref{crossings_table_prague}). 
$\left<C_{true}\right>$ is above 1 in only three languages: Ancient Greek, Latin and Dutch for Stanford dependencies; and only Ancient Greek and Latin for Prague dependencies.
These average numbers of observed crossings are really small when compared against the average potential number of crossings of a linear tree of the same size ($\left<|Q_{linear}|\right>$) or the average potential number of crossings of the same tree ($\left<|Q|\right>$). In order of magnitude, the difference between $\left<|Q_{linear}|\right>$ and $\left<Q\right>$ is small, suggesting that real trees are close to linear trees, namely, their hubiness is low. 

\begin{figure}
\includegraphics[scale = 0.7]{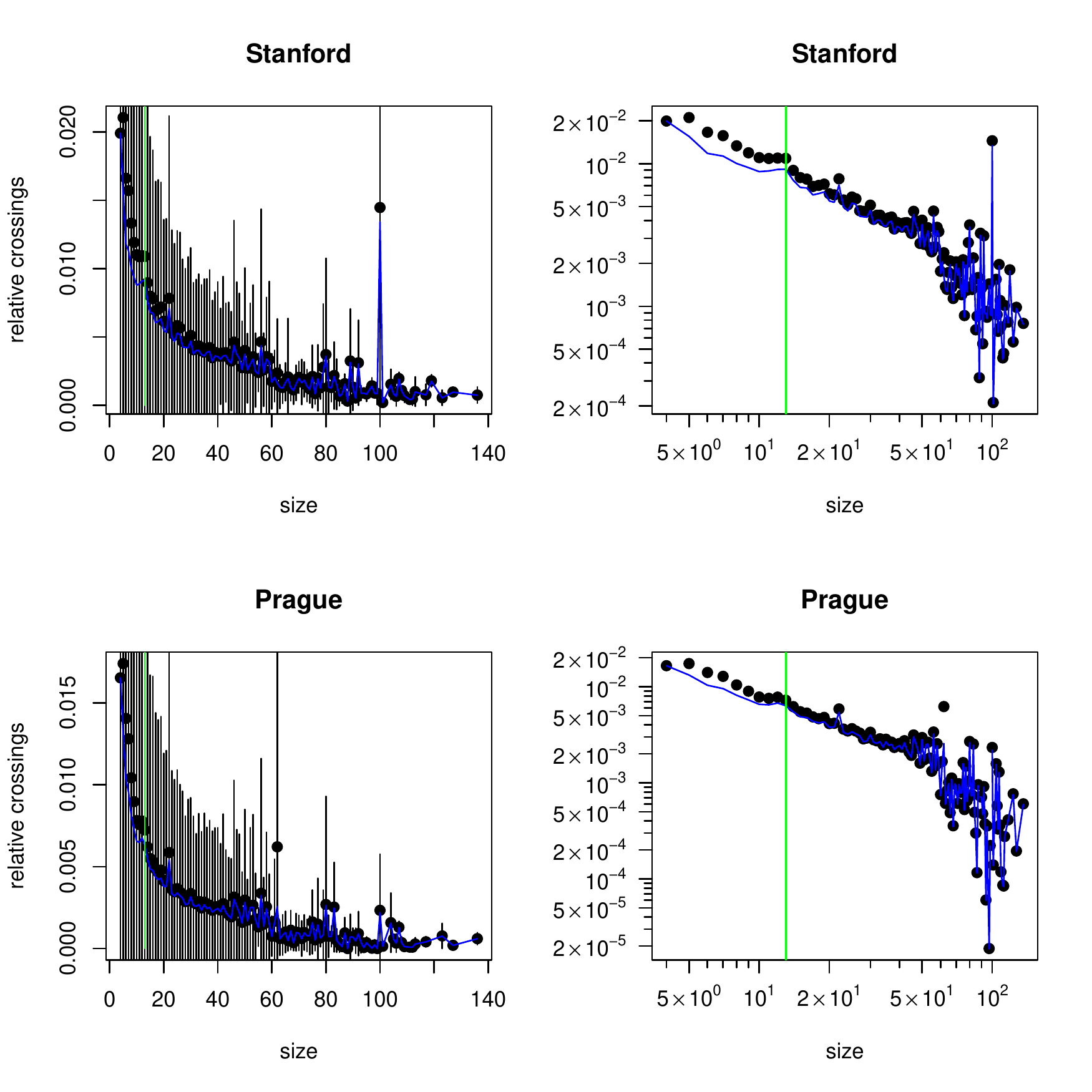}
\caption{\label{relative_crossings_figure} $C_{true}/|Q|$, the relative number of crossings with respect to the potential number of crossings of the tree (black) and $C_{true}/|Q_{linear}|$, the relative number of crossings with respect to the potential number of crossings of a linear tree (blue),
as a function of $n$, the tree size. As a guide to the eye, a vertical line (green) for $n = 13$ is also shown. The format is the same as in Fig. \ref{planarity_figure}. }
\end{figure}

A deeper evaluation of the scarcity of crossing dependencies can be made with the help of ratios between $C_{true}$ and the different baselines: $C_{true}/|Q_{linear}|$, $C_{true}/|Q|$, $C_{true}/E_{URLA}[C]$ and $C_{true}/\doubleexpectationwithoutstars$. 
$C_{true}/|Q|$ has already been used in research on crossings in random trees \cite{Ferrer2014c}. Bear in mind that 
\begin{itemize}
\item
All these ratios are positive but only $C_{true}/|Q_{linear}|$ and $C_{true}/|Q|$ are bounded above by $1$.
\item
Each ratio defined on random baselines is proportional or approximately proportional to a deterministic baseline. On the one hand, 
\begin{equation}
\frac{C_{true}}{E_{URLA}[C]} = 3 \frac{C_{true}}{|Q|} \label{proportionality_equation}
\end{equation} 
thanks to Eq. \ref{expected_number_of_crossings_equation}.
On the other hand, 
\begin{equation}
\frac{C_{true}}{\doubleexpectationwithoutstars} = \frac{3(n-n^{4-n})}{n -1} \frac{C_{true}}{|Q_{linear}|} 
\end{equation} 
thanks to Eqs. \ref{potential_number_of_crossings_of_linear_tree_equation} and 
\ref{main_adapted_expected_number_of_crossings_of_uniformly_random_tree_equation}.
Then 
\begin{equation}
\frac{C_{true}}{\doubleexpectationwithoutstars} \approx 3 \frac{C_{true}}{|Q_{linear}|} \label{proportionality_linear_equation}
\end{equation} 
for sufficiently large $n$.
\item
Although  
\begin{equation}
\frac{C_{true}}{|Q_{linear}|} \leq \frac{C_{true}}{|Q|}
\end{equation}
thanks to $|Q| \leq |Q_{linear}$, the relationship between 
\begin{equation}
\frac{C_{true}}{E_{URLA}[C]}
\end{equation}
and 
\begin{equation}
\frac{C_{true}}{\doubleexpectationwithoutstars}
\end{equation}
is uncertain.
\end{itemize}
$C_{true}/|Q_{linear}|$ and $C_{true}/|Q|$ tend to decrease as tree size increases (Fig. \ref{relative_crossings_figure}) and the same is expected to happen to their corresponding random baselines thanks to the proportionality relationships above (Eqs. \ref{proportionality_equation} and \ref{proportionality_linear_equation}). Therefore, the evidence of the scarcity of crossings increases as tree size increases.

The ratios in Tables \ref{relative_crossings_table_stanford} and \ref{relative_crossings_table_prague} show that, on average, the actual number of crossings is smaller than that of the baseline for all treebanks and for all baselines: all average ratios are below 0.3. These ratios allow one to analyze with more detail the difference in magnitude between $C_{true}$ and the different baselines (Tables \ref{relative_crossings_table_stanford} and \ref{relative_crossings_table_prague}):
\begin{itemize}
\item
$\left<C_{true}/|Q_{linear}|\right>$ indicates that, on average, $C_{true}$ is at least 10 times smaller than $|Q_{linear}|$ and $|Q|$ across languages. The smallest differences are achieved by Ancient Greek, where $\left< C_{true}/|Q_{linear}| \right> \approx 0.07$ and $\left< C_{true}/|Q|\right> \approx 0.09$.
The relative number of crossings with respect to the same tree, i.e., $C_{true}/|Q|$, is expected to be about $1/3$ in a random linear arrangement of vertices \cite{Ferrer2014c} but indeed it is much smaller.
\item
$\left<C_{true}/|Q_{linear}|\right> \leq \left<C_{true}/|Q|\right>$ as expected but the difference between $\left<C_{true}/|Q_{linear}|\right>$ and 
$\left<C_{true}/|Q|\right>$ is small, suggesting that real trees are closer to linear trees than to star trees.  
\item
$\left<C_{true}/\doubleexpectationwithoutstars \right>$ indicates that, on average, $C_{true}$ is at least 10 times smaller than $\doubleexpectationwithoutstars]$ across treebanks except for Ancient Greek and Latin. For Ancient Greek, $\left< C_{true}/\doubleexpectationwithoutstars \right> \approx 0.23$ on average with Stanford dependencies and $\left<C_{true}/\doubleexpectationwithoutstars \right> \approx 0.24$ with Prague dependencies. For Latin, $\left<C_{true}/\doubleexpectationwithoutstars \right> \approx 0.14$ on average with Stanford dependencies and $\left<C_{true}/\doubleexpectationwithoutstars \right> \approx 0.13$ with Prague dependencies. 
\item
$\left< C_{true}/E_{URLA}[C] \right>$ indicates that, on average, $C_{true}$ is at least 10 times smaller than $E_{URLA}[C]$ across treebanks except for Ancient Greek and Latin.
For Ancient Greek, $\left< C_{true}/E_{URLA}[C] \right> \approx 0.27$ on average with both Stanford and Prague dependencies. For Latin, $\left< C_{true}/E_{URLA}[C] \right> \approx 0.15$ with Stanford dependencies and $\left< C_{true}/E_{URLA}[C] \right> \approx 0.14$ with Prague dependencies.
\item
The difference between $\left<C_{true}/\doubleexpectationwithoutstars \right>$ and $\left<C_{true}/E_{URLA}[C]\right>$ is small. The condition $\left<C_{true}/\doubleexpectationwithoutstars \right> \leq \left<C_{true}/E_{URLA}[C]\right>$ holds for all treebanks with Stanford annotations, 
as well as for all treebanks with Prague annotations except for Japanese and Persian. 
\end{itemize}

\changed{The significance of the gap that separates the actual number of crossings and the predictions of random baselines must be evaluated statistically. Indeed, $p(C_{true} = 0)$ and $\left< C_{true} \right>$ are smaller than expected by URLTs and URLAs: the Monte Carlo test described in Section \ref{materials_section} yields left-$P < 10^{-4}$ for all the treebanks and both random baselines.
}

\begin{figure}
\includegraphics[scale = 0.7]{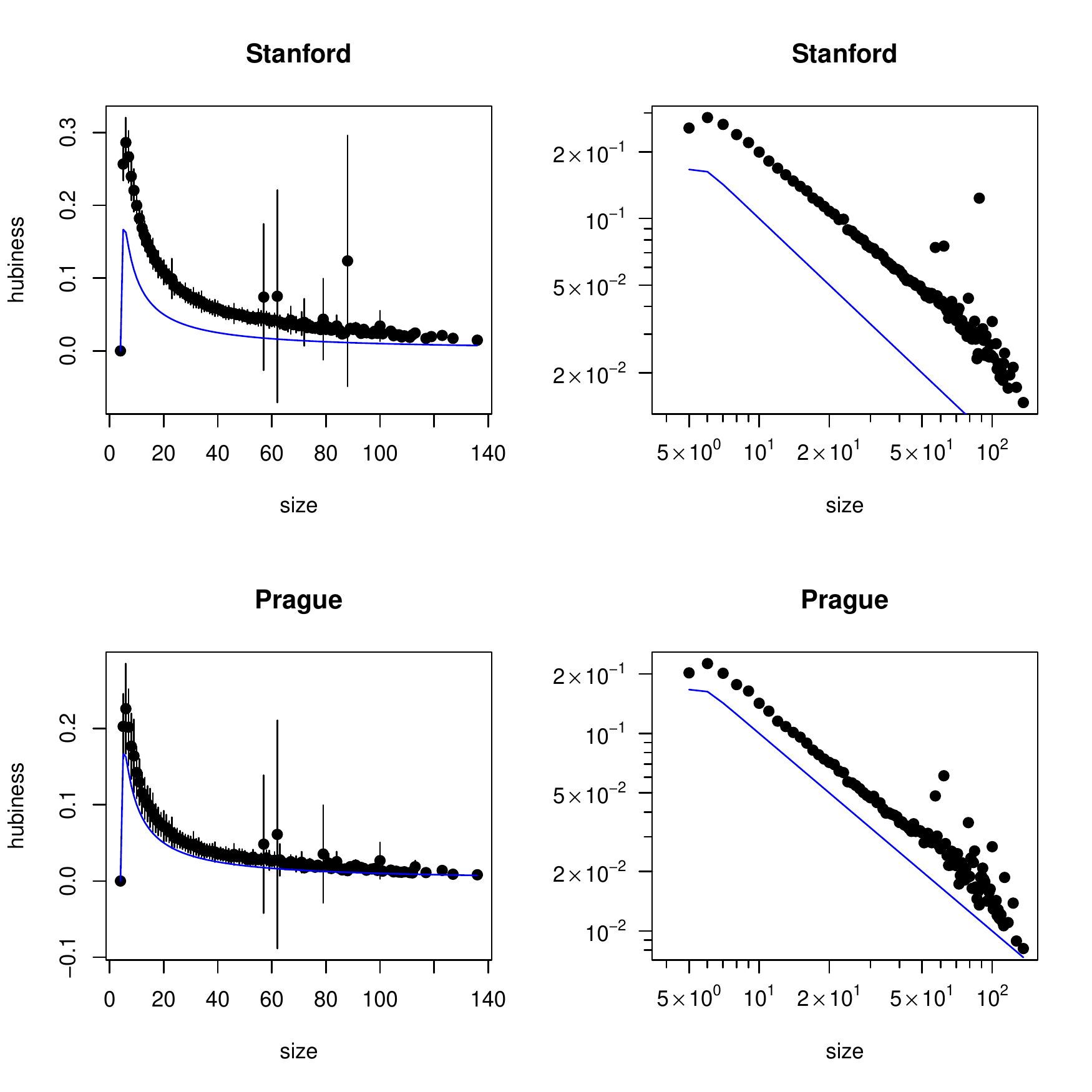}
\caption{\label{hubiness_figure} $h$, the hubiness coefficient (black), and $E_{URLT}[h| \neg \mbox{star}] \approx 1/n$, the expected hubiness coefficient of a uniformly random labelled tree excluding star trees (blue), as a function of $n$, the tree size. The format is the same as in Fig. \ref{planarity_figure}. When $n = 4$ a tree can only be a star tree or a linear tree. As our analysis excludes star trees (Section \ref{materials_section}), $h = 0$ for $n = 4$. For this reason $n = 4$ is included in normal scale but excluded in log-log scale.   
}
\end{figure}

Fig. \ref{hubiness_figure} shows that the hubiness of trees tends to decrease as $n$ increases.  
Tables \ref{hubiness_table_stanford} and \ref{hubiness_table_prague} also show that $\left<h\right>$ never exceeds $0.24$ and is $\approx 0.1$ across treebanks, suggesting that real trees are closer to linear trees than to star trees. The similarity between linear trees and real trees supports the little difference reported above between $\left<C_{true}/|Q_{linear}|\right>$ and $\left<C_{true}/Q\right>$. Indeed, recall the alternative definition of $h$ in Eq. \ref{hubiness_vs_potential_crossings_equation}.
Concerning URLTs, Fig. \ref{hubiness_figure} shows that the average hubiness of real sentences tends to be above the average hubiness that is expected in a URLT over the ensemble of treebanks. A detailed analysis reveals that the average hubiness of real sentences is above the average hubiness that is expected in a URLT for all treebanks with Stanford dependencies (Table \ref{hubiness_table_stanford}). However, this does not hold for the \changed{Arabic}, Japanese and Persian 
treebank with Prague dependencies (Table \ref{hubiness_table_prague}) but the difference is small. The systematic deviation between $\left<h\right>$ and $\left< E_{URLT}[h | \neg \mbox{star}] \right>$ suggests that the hubiness of real dependency trees cannot be explained by sampling of URLTs, especially for Stanford dependencies. The gap between URLTs and real syntactic dependency trees is smaller for Prague dependencies, as Fig. \ref{hubiness_figure} suggests. Notice that 
$\left<h\right>$ is about twice $\left< E_{URLT}[h | \neg \mbox{star}] \right>$ with Stanford dependencies whereas $\left<h\right>$ is about 1.4 times $\left< E_{URLT}[h | \neg \mbox{star}] \right>$ with Prague dependencies. \changed{The Monte Carlo tests indicate that $\left<h\right>$ is significantly large in all treebanks with Stanford annotations (right-$P < 10^{-4}$). The results are less homogeneous for Prague annotations: $\left<h\right>$ is significantly small in Arabic, Japanese and Persian (left-$P < 10^{-4}$) but significantly large for the remainder (right-$P < 10^{-4}$ in all cases except right-$P = 10^{-4}$ for Portuguese). }
 
\begin{table}
\caption{\label{crossings_table_stanford} A summary of the analysis of the number of crossings in treebanks from different languages, annotated under the Stanford guidelines. First, $p(C_{true} = 0)$, the proportion of planar sentences of the treebanks. Second, mean and standard deviation over the whole treebank for a series of measures: $C_{true}$, the actual number of crossings, $|Q_{linear}|$, the maximum potential number of crossings and $|Q|$, the potential number of crossings. For every metric (except $p(C_{true} = 0)$), we show $\mu \pm \sigma$, where $\mu$ is the average value of the measures over all the trees of the treebank included in our analysis and $\sigma$ is their standard deviation.    
}
\begin{scriptsize}
\begin{center}
\begin{tabular}{lrrrr}
\hline
Language & $p(C_{true} = 0)$ & $C_{true}$ & $|Q_{linear}|$ & $|Q|$ \\ \hline
\textbf{Arabic} & 0.690 & $ 0.981 \pm 2.059 $ & $ 505.0 \pm 967.9 $ & $ 486.2 \pm 949.1 $  \\ 
\textbf{Basque} & 0.932 & $ 0.139 \pm 0.684 $ & $ 56.4 \pm 63.8 $ & $ 48.3 \pm 58.2 $  \\ 
\textbf{Bengali} & 0.944 & $ 0.106 \pm 0.627 $ & $ 16.9 \pm 24.4 $ & $ 13.4 \pm 21.4 $  \\ 
\textbf{Bulgarian} & 0.843 & $ 0.360 \pm 1.024 $ & $ 90.1 \pm 150.0 $ & $ 80.9 \pm 143.3 $  \\ 
\textbf{Catalan} & 0.790 & $ 0.642 \pm 1.606 $ & $ 394.0 \pm 476.7 $ & $ 366.9 \pm 460.1 $  \\ 
\textbf{Czech} & 0.780 & $ 0.528 \pm 1.277 $ & $ 135.7 \pm 193.3 $ & $ 124.0 \pm 184.5 $  \\ 
\textbf{Danish} & 0.734 & $ 0.680 \pm 1.482 $ & $ 155.6 \pm 223.4 $ & $ 141.6 \pm 212.8 $  \\ 
\textbf{Dutch} & 0.654 & $ 1.398 \pm 2.690 $ & $ 95.6 \pm 148.1 $ & $ 86.0 \pm 139.9 $  \\ 
\textbf{English} & 0.787 & $ 0.524 \pm 1.370 $ & $ 233.8 \pm 243.1 $ & $ 213.7 \pm 231.0 $  \\  
\textbf{Estonian} & 0.974 & $ 0.038 \pm 0.267 $ & $ 19.6 \pm 41.3 $ & $ 15.9 \pm 36.8 $  \\  
\textbf{Finnish} & 0.891 & $ 0.318 \pm 1.215 $ & $ 62.4 \pm 79.0 $ & $ 54.6 \pm 73.4 $  \\ 
\textbf{German} & 0.684 & $ 0.783 \pm 1.593 $ & $ 150.9 \pm 229.4 $ & $ 138.2 \pm 219.9 $  \\ 
\textbf{Greek (Anc.)} & 0.312 & $ 3.262 \pm 4.880 $ & $ 89.1 \pm 176.0 $ & $ 77.0 \pm 147.1 $  \\ 
\textbf{Greek (Mod.)} & 0.752 & $ 0.654 \pm 1.513 $ & $ 289.2 \pm 374.2 $ & $ 269.7 \pm 361.0 $  \\ 
\textbf{Hindi} & 0.858 & $ 0.304 \pm 0.988 $ & $ 202.8 \pm 222.5 $ & $ 183.3 \pm 211.0 $  \\ 
\textbf{Hungarian} & 0.728 & $ 0.972 \pm 2.426 $ & $ 190.8 \pm 245.4 $ & $ 173.7 \pm 232.5 $  \\ 
\textbf{Italian} & 0.851 & $ 0.415 \pm 1.309 $ & $ 199.0 \pm 341.3 $ & $ 183.9 \pm 329.2 $  \\ 
\textbf{Japanese} & 0.884 & $ 0.164 \pm 0.523 $ & $ 51.2 \pm 95.3 $ & $ 46.3 \pm 90.6 $  \\ 
\textbf{Latin} & 0.505 & $ 2.179 \pm 3.920 $ & $ 115.1 \pm 179.8 $ & $ 102.7 \pm 168.3 $  \\ 
\textbf{Persian} & 0.785 & $ 0.591 \pm 2.833 $ & $ 130.8 \pm 361.8 $ & $ 121.2 \pm 351.9 $  \\
\textbf{Portuguese} & 0.751 & $ 0.634 \pm 1.452 $ & $ 252.8 \pm 401.2 $ & $ 236.2 \pm 388.8 $  \\ 
\textbf{Romanian} & 0.953 & $ 0.102 \pm 0.474 $ & $ 52.6 \pm 96.5 $ & $ 46.8 \pm 91.1 $  \\ 
\textbf{Russian} & 0.810 & $ 0.417 \pm 1.131 $ & $ 119.8 \pm 194.0 $ & $ 109.5 \pm 185.9 $  \\ 
\textbf{Slovak} & 0.819 & $ 0.456 \pm 1.267 $ & $ 109.9 \pm 234.2 $ & $ 99.3 \pm 225.4 $  \\ 
\textbf{Slovenian} & 0.775 & $ 0.775 \pm 2.068 $ & $ 143.4 \pm 257.1 $ & $ 127.6 \pm 244.2 $  \\
\textbf{Spanish} & 0.794 & $ 0.622 \pm 1.594 $ & $ 397.4 \pm 432.6 $ & $ 371.2 \pm 416.8 $  \\ 
\textbf{Swedish} & 0.824 & $ 0.487 \pm 1.764 $ & $ 133.2 \pm 209.8 $ & $ 120.7 \pm 199.6 $  \\ 
\textbf{Tamil} & 0.979 & $ 0.024 \pm 0.174 $ & $ 100.3 \pm 141.8 $ & $ 89.6 \pm 133.3 $  \\ 
\textbf{Telugu} & 0.986 & $ 0.014 \pm 0.117 $ & $ 5.5 \pm 7.7 $ & $ 4.3 \pm 6.3 $  \\ 
\textbf{Turkish} & 0.945 & $ 0.098 \pm 0.524 $ & $ 74.5 \pm 134.4 $ & $ 66.9 \pm 126.8 $  \\ \hline
\textbf{Macro avg} & 0.800 & $ 0.622 \pm 1.495 $ & $ 152.4 \pm 231.5 $ & $ 140.0 \pm 221.3 $  \\ \hline
\end{tabular}
\end{center}
\end{scriptsize}
\end{table}

\begin{table}
\caption{\label{crossings_table_prague} A summary of the analysis of the number of crossings in treebanks from different languages, annotated under the Prague guidelines. The format and data shown are as in Table \ref{crossings_table_stanford}.  
}
\begin{scriptsize}
\begin{center}
\begin{tabular}{lrrrr}
\hline
Language & $p(C_{true} = 0)$ & $C_{true}$ & $|Q_{linear}|$ & $|Q|$ \\ \hline
\textbf{Arabic} & 0.945 & $ 0.088 \pm 0.458 $ & $ 507.6 \pm 968.3 $ & $ 496.0 \pm 956.1 $  \\ 
\textbf{Basque} & 0.933 & $ 0.125 \pm 0.641 $ & $ 55.6 \pm 63.3 $ & $ 49.3 \pm 59.3 $  \\ 
\textbf{Bengali} & 0.939 & $ 0.124 \pm 0.693 $ & $ 15.9 \pm 21.4 $ & $ 12.6 \pm 18.8 $  \\ 
\textbf{Bulgarian} & 0.905 & $ 0.125 \pm 0.426 $ & $ 88.3 \pm 147.5 $ & $ 83.0 \pm 143.7 $  \\ 
\textbf{Catalan} & 0.955 & $ 0.087 \pm 0.532 $ & $ 392.9 \pm 476.4 $ & $ 376.4 \pm 466.9 $  \\ 
\textbf{Czech} & 0.785 & $ 0.373 \pm 0.920 $ & $ 132.2 \pm 191.4 $ & $ 124.7 \pm 186.0 $  \\ 
\textbf{Danish} & 0.880 & $ 0.164 \pm 0.519 $ & $ 154.3 \pm 222.2 $ & $ 146.3 \pm 216.1 $  \\ 
\textbf{Dutch} & 0.673 & $ 0.990 \pm 1.928 $ & $ 94.1 \pm 147.4 $ & $ 88.6 \pm 142.9 $  \\ 
\textbf{English} & 0.941 & $ 0.107 \pm 0.811 $ & $ 233.2 \pm 248.3 $ & $ 220.4 \pm 241.2 $  \\  
\textbf{Estonian} & 0.992 & $ 0.013 \pm 0.157 $ & $ 19.0 \pm 41.0 $ & $ 16.0 \pm 37.9 $  \\  
\textbf{Finnish} & 0.908 & $ 0.128 \pm 0.476 $ & $ 61.6 \pm 78.8 $ & $ 56.1 \pm 75.2 $  \\ 
\textbf{German} & 0.671 & $ 0.723 \pm 1.489 $ & $ 148.3 \pm 227.9 $ & $ 140.0 \pm 221.8 $  \\ 
\textbf{Greek (Anc.)} & 0.323 & $ 3.353 \pm 4.446 $ & $ 89.5 \pm 181.7 $ & $ 79.1 \pm 151.6 $  \\ 
\textbf{Greek (Mod.)} & 0.867 & $ 0.206 \pm 0.660 $ & $ 286.6 \pm 368.4 $ & $ 274.2 \pm 360.1 $  \\ 
\textbf{Hindi} & 0.769 & $ 0.387 \pm 0.958 $ & $ 201.8 \pm 221.7 $ & $ 189.4 \pm 215.0 $  \\ 
\textbf{Hungarian} & 0.738 & $ 0.867 \pm 2.143 $ & $ 185.9 \pm 247.8 $ & $ 172.3 \pm 237.6 $  \\ 
\textbf{Italian} & 0.959 & $ 0.062 \pm 0.364 $ & $ 196.8 \pm 343.3 $ & $ 187.5 \pm 335.7 $  \\ 
\textbf{Japanese} & 1.000 & $ 0.000 \pm 0.014 $ & $ 49.4 \pm 93.9 $ & $ 46.8 \pm 91.3 $  \\ 
\textbf{Latin} & 0.499 & $ 1.850 \pm 3.171 $ & $ 114.8 \pm 180.2 $ & $ 106.2 \pm 172.5 $  \\ 
\textbf{Persian} & 0.817 & $ 0.402 \pm 2.122 $ & $ 125.3 \pm 355.5 $ & $ 120.2 \pm 349.8 $  \\
\textbf{Portuguese} & 0.860 & $ 0.247 \pm 0.780 $ & $ 250.6 \pm 400.9 $ & $ 241.6 \pm 394.1 $  \\ 
\textbf{Romanian} & 1.000 & $ 0.000 \pm 0.000 $ & $ 51.8 \pm 95.9 $ & $ 48.0 \pm 92.5 $  \\ 
\textbf{Russian} & 0.907 & $ 0.157 \pm 0.585 $ & $ 118.9 \pm 193.7 $ & $ 112.6 \pm 188.8 $  \\ 
\textbf{Slovak} & 0.853 & $ 0.269 \pm 0.829 $ & $ 102.9 \pm 197.6 $ & $ 96.3 \pm 191.3 $  \\ 
\textbf{Slovenian} & 0.822 & $ 0.312 \pm 0.802 $ & $ 128.1 \pm 224.0 $ & $ 118.4 \pm 216.4 $  \\ 
\textbf{Spanish} & 0.945 & $ 0.110 \pm 0.625 $ & $ 395.4 \pm 432.7 $ & $ 380.2 \pm 423.8 $  \\
\textbf{Swedish} & 0.935 & $ 0.195 \pm 1.476 $ & $ 129.1 \pm 202.8 $ & $ 120.0 \pm 195.6 $  \\ 
\textbf{Tamil} & 0.988 & $ 0.014 \pm 0.130 $ & $ 100.1 \pm 141.8 $ & $ 92.2 \pm 135.8 $  \\ 
\textbf{Telugu} & 0.992 & $ 0.008 \pm 0.089 $ & $ 5.3 \pm 7.5 $ & $ 4.1 \pm 6.2 $  \\ 
\textbf{Turkish} & 0.914 & $ 0.137 \pm 0.540 $ & $ 67.1 \pm 123.4 $ & $ 63.2 \pm 119.6 $  \\ \hline
\textbf{Macro avg} & 0.857 & $ 0.388 \pm 0.959 $ & $ 150.1 \pm 228.2 $ & $ 142.0 \pm 221.5 $  \\ \hline
\end{tabular}
\end{center}
\end{scriptsize}
\end{table}

\begin{table}
\caption{\label{relative_crossings_table_stanford} A summary of the normalized number of crossings in treebanks from different languages (Stanford annotation): $C_{true}/|Q_{linear}|$, $C_{true}/|Q|$, 
 $C_{true}/\doubleexpectationwithoutstars$, $C_{true}/E_{URLA}[C]$. The format is as in Table \ref{crossings_table_stanford}.}
\begin{scriptsize}
\begin{center}
\begin{tabular}{lrrrr}
\hline
Language & $\frac{C_{true}}{|Q_{linear}|}$ & $\frac{C_{true}}{|Q|}$ & $\frac{C_{true}}{\doubleexpectationwithoutstars}$ & $\frac{C_{true}}{E_{URLA}[C]}$ \\ \hline
\textbf{Arabic} & $ 0.00464 \pm 0.01769 $ & $ 0.00534 \pm 0.02118 $      & $ 0.01523 \pm 0.06006 $ & $ 0.01602 \pm 0.06353 $ \\
\textbf{Basque} & $ 0.00300 \pm 0.02141 $ & $ 0.00357 \pm 0.02519 $      & $ 0.01006 \pm 0.07237 $ & $ 0.01070 \pm 0.07558 $ \\
\textbf{Bengali} & $ 0.01023 \pm 0.07418 $ & $ 0.01237 \pm 0.08446 $     & $ 0.03350 \pm 0.23144 $ & $ 0.03711 \pm 0.25339 $ \\
\textbf{Bulgarian} & $ 0.00566 \pm 0.02778 $ & $ 0.00685 \pm 0.03605 $   & $ 0.01883 \pm 0.09411 $ & $ 0.02054 \pm 0.10816 $ \\
\textbf{Catalan} & $ 0.00231 \pm 0.01062 $ & $ 0.00267 \pm 0.01463 $     & $ 0.00739 \pm 0.03577 $ & $ 0.00800 \pm 0.04388 $ \\
\textbf{Czech} & $ 0.00644 \pm 0.03073 $ & $ 0.00767 \pm 0.03880 $       & $ 0.02123 \pm 0.10251 $ & $ 0.02301 \pm 0.11639 $ \\
\textbf{Danish} & $ 0.00731 \pm 0.03158 $ & $ 0.00924 \pm 0.04160 $      & $ 0.02398 \pm 0.10346 $ & $ 0.02771 \pm 0.12481 $ \\
\textbf{Dutch} & $ 0.01867 \pm 0.05240 $ & $ 0.02274 \pm 0.07002 $       & $ 0.06225 \pm 0.18013 $ & $ 0.06821 \pm 0.21005 $ \\
\textbf{English} & $ 0.00263 \pm 0.00983 $ & $ 0.00301 \pm 0.01226 $     & $ 0.00839 \pm 0.03256 $ & $ 0.00904 \pm 0.03677 $ \\
\textbf{Estonian} & $ 0.00213 \pm 0.01835 $ & $ 0.00276 \pm 0.02492 $    & $ 0.00743 \pm 0.06480 $ & $ 0.00827 \pm 0.07476 $ \\
\textbf{Finnish} & $ 0.00615 \pm 0.02993 $ & $ 0.00756 \pm 0.04028 $     & $ 0.02056 \pm 0.10358 $ & $ 0.02269 \pm 0.12084 $ \\
\textbf{German} & $ 0.00745 \pm 0.03075 $ & $ 0.00870 \pm 0.03777 $      & $ 0.02440 \pm 0.10243 $ & $ 0.02611 \pm 0.11331 $ \\
\textbf{Greek (Anc.)} & $ 0.06789 \pm 0.11445 $ & $ 0.08954 \pm 0.15792 $  & $ 0.22852 \pm 0.38975 $ & $ 0.26861 \pm 0.47376 $ \\
\textbf{Greek (Mod.)} & $ 0.00303 \pm 0.00991 $ & $ 0.00343 \pm 0.01171 $       & $ 0.00969 \pm 0.03249 $ & $ 0.01029 \pm 0.03512 $ \\
\textbf{Hindi} & $ 0.00148 \pm 0.00650 $ & $ 0.00174 \pm 0.00943 $       & $ 0.00472 \pm 0.02186 $ & $ 0.00522 \pm 0.02828 $ \\
\textbf{Hungarian} & $ 0.00651 \pm 0.02523 $ & $ 0.00768 \pm 0.03247 $   & $ 0.02107 \pm 0.08304 $ & $ 0.02305 \pm 0.09741 $ \\
\textbf{Italian} & $ 0.00354 \pm 0.02341 $ & $ 0.00429 \pm 0.03274 $     & $ 0.01172 \pm 0.08253 $ & $ 0.01288 \pm 0.09822 $ \\
\textbf{Japanese} & $ 0.00674 \pm 0.05244 $ & $ 0.00760 \pm 0.05628 $    & $ 0.02213 \pm 0.16395 $ & $ 0.02279 \pm 0.16883 $ \\
\textbf{Latin} & $ 0.04057 \pm 0.09173 $ & $ 0.05094 \pm 0.12026 $       & $ 0.13626 \pm 0.31224 $ & $ 0.15281 \pm 0.36077 $ \\
\textbf{Persian} & $ 0.00639 \pm 0.03244 $ & $ 0.00765 \pm 0.04100 $     & $ 0.02116 \pm 0.10850 $ & $ 0.02294 \pm 0.12300 $ \\
\textbf{Portuguese} & $ 0.00438 \pm 0.02073 $ & $ 0.00506 \pm 0.02530 $  & $ 0.01433 \pm 0.06949 $ & $ 0.01519 \pm 0.07589 $ \\
\textbf{Romanian} & $ 0.00182 \pm 0.01237 $ & $ 0.00208 \pm 0.01443 $    & $ 0.00606 \pm 0.04228 $ & $ 0.00625 \pm 0.04330 $ \\
\textbf{Russian} & $ 0.00657 \pm 0.03949 $ & $ 0.00782 \pm 0.04703 $     & $ 0.02165 \pm 0.12800 $ & $ 0.02347 \pm 0.14108 $ \\
\textbf{Slovak} & $ 0.00787 \pm 0.04410 $ & $ 0.00959 \pm 0.05430 $      & $ 0.02617 \pm 0.14497 $ & $ 0.02878 \pm 0.16291 $ \\
\textbf{Slovenian} & $ 0.00901 \pm 0.04283 $ & $ 0.01112 \pm 0.05502 $   & $ 0.02984 \pm 0.14335 $ & $ 0.03335 \pm 0.16507 $ \\
\textbf{Spanish} & $ 0.00233 \pm 0.01151 $ & $ 0.00266 \pm 0.01462 $     & $ 0.00745 \pm 0.03934 $ & $ 0.00797 \pm 0.04384 $ \\
\textbf{Swedish} & $ 0.00485 \pm 0.02292 $ & $ 0.00569 \pm 0.02809 $     & $ 0.01591 \pm 0.07725 $ & $ 0.01708 \pm 0.08427 $ \\
\textbf{Tamil} & $ 0.00044 \pm 0.00365 $ & $ 0.00053 \pm 0.00437 $       & $ 0.00145 \pm 0.01211 $ & $ 0.00158 \pm 0.01310 $ \\
\textbf{Telugu} & $ 0.01049 \pm 0.09766 $ & $ 0.01088 \pm 0.09861 $      & $ 0.03216 \pm 0.29508 $ & $ 0.03263 \pm 0.29583 $ \\ 
\textbf{Turkish} & $ 0.00298 \pm 0.03280 $ & $ 0.00363 \pm 0.03662 $     & $ 0.00987 \pm 0.10351 $ & $ 0.01090 \pm 0.10985 $ \\ \hline
\textbf{Macro avg} & $ 0.00878 \pm 0.03465 $ & $ 0.01081 \pm 0.04291 $   & $ 0.02911 \pm 0.11443 $ & $ 0.03244 \pm 0.12873 $ \\ \hline
\end{tabular}
\end{center}
\end{scriptsize}
\end{table}

\begin{table}
\caption{\label{relative_crossings_table_prague} A summary of the normalized number of crossings in treebanks from different languages (Prague annotation). The format and data shown are as in Table \ref{relative_crossings_table_stanford}.}
\begin{scriptsize}
\begin{center}
\begin{tabular}{lrrrr}
\hline
Language & $\frac{C_{true}}{|Q_{linear}|}$ & $\frac{C_{true}}{|Q|}$ & $\frac{C_{true}}{\doubleexpectationwithoutstars}$ & $\frac{C_{true}}{E_{URLA}[C]}$ \\ \hline
\textbf{Arabic} & $ 0.00024 \pm 0.00317 $ & $ 0.00026 \pm 0.00362 $ &                          $ 0.00077              \pm 0.01022 $ & $ 0.00078              \pm 0.01085 $ \\
\textbf{Basque} & $ 0.00282 \pm 0.02033 $ & $ 0.00328 \pm 0.02366 $ &                          $ 0.00945              \pm 0.06834 $ & $ 0.00984              \pm 0.07097 $ \\
\textbf{Bengali} & $ 0.01024 \pm 0.07027 $ & $ 0.01291 \pm 0.08603 $ &                         $ 0.03417              \pm 0.22580 $ & $ 0.03873              \pm 0.25808 $ \\ 
\textbf{Bulgarian} & $ 0.00364 \pm 0.03441 $ & $ 0.00407 \pm 0.03688 $ &                       $ 0.01208              \pm 0.10910 $ & $ 0.01221              \pm 0.11063 $ \\
\textbf{Catalan} & $ 0.00024 \pm 0.00226 $ & $ 0.00027 \pm 0.00269 $ &                         $ 0.00077              \pm 0.00743 $ & $ 0.00081              \pm 0.00806 $ \\
\textbf{Czech} & $ 0.00536 \pm 0.02856 $ & $ 0.00605 \pm 0.03273 $ &                           $ 0.01774              \pm 0.09367 $ & $ 0.01816              \pm 0.09818 $ \\
\textbf{Danish} & $ 0.00164 \pm 0.00817 $ & $ 0.00192 \pm 0.01094 $ &                          $ 0.00538              \pm 0.02749 $ & $ 0.00575              \pm 0.03281 $ \\
\textbf{Dutch} & $ 0.01585 \pm 0.05088 $ & $ 0.01858 \pm 0.06694 $ &                           $ 0.05325              \pm 0.17535 $ & $ 0.05573              \pm 0.20081 $ \\
\textbf{English} & $ 0.00060 \pm 0.00594 $ & $ 0.00066 \pm 0.00697 $ &                         $ 0.00194              \pm 0.01991 $ & $ 0.00199              \pm 0.02090 $ \\
\textbf{Estonian} & $ 0.00077 \pm 0.01239 $ & $ 0.00114 \pm 0.01914 $ &                        $ 0.00268              \pm 0.04353 $ & $ 0.00343              \pm 0.05742 $ \\
\textbf{Finnish} & $ 0.00254 \pm 0.01391 $ & $ 0.00300 \pm 0.01759 $ &                         $ 0.00850              \pm 0.04832 $ & $ 0.00901              \pm 0.05277 $ \\
\textbf{German} & $ 0.00659 \pm 0.02447 $ & $ 0.00738 \pm 0.02824 $ &                          $ 0.02157              \pm 0.08079 $ & $ 0.02213              \pm 0.08473 $ \\
\textbf{Greek (Anc.)} & $ 0.07030 \pm 0.11966 $ & $ 0.09050 \pm 0.16363 $ &                      $ 0.23676              \pm 0.40801 $ & $ 0.27151              \pm 0.49088 $ \\
\textbf{Greek (Mod.)} & $ 0.00103 \pm 0.00515 $ & $ 0.00112 \pm 0.00574 $ &                           $ 0.00331              \pm 0.01716 $ & $ 0.00337              \pm 0.01723 $ \\
\textbf{Hindi} & $ 0.00261 \pm 0.00915 $ & $ 0.00295 \pm 0.01148 $ &                           $ 0.00846              \pm 0.03122 $ & $ 0.00885              \pm 0.03443 $ \\
\textbf{Hungarian} & $ 0.00597 \pm 0.02448 $ & $ 0.00691 \pm 0.02964 $ &                       $ 0.01930              \pm 0.07972 $ & $ 0.02075              \pm 0.08891 $ \\
\textbf{Italian} & $ 0.00046 \pm 0.00461 $ & $ 0.00053 \pm 0.00587 $ &                         $ 0.00149              \pm 0.01538 $ & $ 0.00159              \pm 0.01762 $ \\
\textbf{Japanese} & \tiny{$\expnumber{7.45}{-6} \pm 0.00052$} & \tiny{$\expnumber{7.73}{-6} \pm 0.00053 $} & \tiny{$ \expnumber{2.48}{-5} \pm 0.00172 $} & \tiny{$ \expnumber{2.32}{-5} \pm 0.00160 $} \\ 
\textbf{Latin} & $ 0.03938 \pm 0.09014 $ & $ 0.04750 \pm 0.11558 $ &                           $ 0.13298              \pm 0.30915 $ & $ 0.14250              \pm 0.34675 $ \\
\textbf{Persian} & $ 0.00436 \pm 0.02215 $ & $ 0.00469 \pm 0.02414 $ &                         $ 0.01444              \pm 0.07363 $ & $ 0.01408              \pm 0.07242 $ \\
\textbf{Portuguese} & $ 0.00181 \pm 0.01537 $ & $ 0.00200 \pm 0.01733 $ &                      $ 0.00592              \pm 0.05020 $ & $ 0.00601              \pm 0.05199 $ \\
\textbf{Romanian} & $ 0.00000 \pm 0.00000 $ & $ 0.00000 \pm 0.00000 $ &                        $ 0.00000              \pm 0.00000 $ & $ 0.00000              \pm 0.00000 $ \\
\textbf{Russian} & $ 0.00308 \pm 0.02864 $ & $ 0.00349 \pm 0.03244 $ &                         $ 0.01017              \pm 0.09206 $ & $ 0.01046              \pm 0.09731 $ \\
\textbf{Slovak} & $ 0.00632 \pm 0.04273 $ & $ 0.00740 \pm 0.05023 $ &                          $ 0.02111              \pm 0.13945 $ & $ 0.02220              \pm 0.15069 $ \\
\textbf{Slovenian} & $ 0.00531 \pm 0.03492 $ & $ 0.00639 \pm 0.04142 $ &                       $ 0.01771              \pm 0.11419 $ & $ 0.01918              \pm 0.12426 $ \\
\textbf{Spanish} & $ 0.00039 \pm 0.00376 $ & $ 0.00049 \pm 0.00632 $ &                         $ 0.00125              \pm 0.01222 $ & $ 0.00146              \pm 0.01896 $ \\
\textbf{Swedish} & $ 0.00146 \pm 0.01263 $ & $ 0.00168 \pm 0.01503 $ &                         $ 0.00477              \pm 0.04341 $ & $ 0.00504              \pm 0.04510 $ \\
\textbf{Tamil} & $ 0.00036 \pm 0.00351 $ & $ 0.00042 \pm 0.00412 $ &                           $ 0.00118              \pm 0.01168 $ & $ 0.00125              \pm 0.01237 $ \\
\textbf{Telugu} & $ 0.00804 \pm 0.08932 $ & $ 0.00804 \pm 0.08932 $ &                          $ 0.02413              \pm 0.26796 $ & $ 0.02413              \pm 0.26796 $ \\
\textbf{Turkish} & $ 0.00571 \pm 0.04363 $ & $ 0.00640 \pm 0.04716 $ &                         $ 0.01923              \pm 0.14104 $ & $ 0.01921              \pm 0.14148 $ \\ \hline
\textbf{Macro avg} & $ 0.00714 \pm 0.02750 $ & $ 0.00862 \pm 0.03318 $ &                       $ 0.02381              \pm 0.09061 $ & $ 0.02587              \pm 0.09954 $ \\ \hline
\end{tabular}
\end{center}
\end{scriptsize}
\end{table}

\begin{table}
\caption{\label{hubiness_table_stanford} A summary of the hubiness in treebanks from different languages (Stanford annotation): $n$, the size of the tree, $\left<k^2 \right>$, the 2nd moment of degree about zero, $h$, the hubiness coefficient and $E_{URLT}[h | \neg \mbox{star}]$, the expected $h$ in a uniformly random labelled tree excluding star trees, and $p(star)$ the proportion of trees of the treebank that are star trees. The format is as in Table \ref{crossings_table_stanford}. }
\begin{scriptsize}
\begin{center}
\begin{tabular}{lrrrr}
\hline
Language & $n$ & $h$ & $E_{URLT}[h | \neg \mbox{star}]$ & $p(star)$ \\ \hline
\textbf{Arabic} & $ 26.3 \pm 21.0 $ & $ 0.089 \pm 0.076 $ &     $ 0.064 \pm 0.045 $ & 0.005 \\
\textbf{Basque} & $ 11.8 \pm 5.2 $ & $ 0.191 \pm 0.120 $ &      $ 0.095 \pm 0.040 $ & 0.044 \\
\textbf{Bengali} & $ 7.5 \pm 3.0 $ & $ 0.239 \pm 0.160 $ &      $ 0.121 \pm 0.049 $ & 0.163 \\ 
\textbf{Bulgarian} & $ 13.6 \pm 7.6 $ & $ 0.167 \pm 0.114 $ &   $ 0.089 \pm 0.044 $ & 0.033 \\ 
\textbf{Catalan} & $ 26.9 \pm 13.8 $ & $ 0.113 \pm 0.082 $ &    $ 0.049 \pm 0.030 $ & 0.009 \\ 
\textbf{Czech} & $ 16.5 \pm 8.8 $ & $ 0.135 \pm 0.092 $ &       $ 0.075 \pm 0.040 $ & 0.018 \\ 
\textbf{Danish} & $ 17.2 \pm 9.8 $ & $ 0.155 \pm 0.109 $ &      $ 0.074 \pm 0.040 $ & 0.032 \\  
\textbf{Dutch} & $ 13.6 \pm 8.3 $ & $ 0.175 \pm 0.124 $ &       $ 0.090 \pm 0.046 $ & 0.033 \\
\textbf{English} & $ 21.7 \pm 9.9 $ & $ 0.127 \pm 0.084 $ &     $ 0.057 \pm 0.031 $ & 0.011 \\  
\textbf{Estonian} & $ 7.3 \pm 4.0 $ & $ 0.231 \pm 0.169 $ &     $ 0.111 \pm 0.061 $ & 0.237 \\   
\textbf{Finnish} & $ 12.2 \pm 5.6 $ & $ 0.173 \pm 0.113 $ &     $ 0.093 \pm 0.039 $ & 0.026 \\ 
\textbf{German} & $ 17.2 \pm 9.3 $ & $ 0.131 \pm 0.089 $ &      $ 0.072 \pm 0.039 $ & 0.034 \\ 
\textbf{Greek (Anc.)} & $ 13.3 \pm 7.8 $ & $ 0.216 \pm 0.138 $ &  $ 0.093 \pm 0.044 $ & 0.043 \\ 
\textbf{Greek (Mod.)} & $ 23.0 \pm 12.6 $ & $ 0.116 \pm 0.081 $ &      $ 0.057 \pm 0.034 $ & 0.008 \\
\textbf{Hindi} & $ 20.4 \pm 9.2 $ & $ 0.140 \pm 0.083 $ &       $ 0.060 \pm 0.029 $ & 0.004 \\ 
\textbf{Hungarian} & $ 19.0 \pm 10.4 $ & $ 0.141 \pm 0.098 $ &  $ 0.068 \pm 0.038 $ & 0.013 \\ 
\textbf{Italian} & $ 18.6 \pm 11.8 $ & $ 0.144 \pm 0.106 $ &    $ 0.071 \pm 0.042 $ & 0.025 \\ 
\textbf{Japanese} & $ 10.5 \pm 6.2 $ & $ 0.155 \pm 0.121 $ &    $ 0.102 \pm 0.050 $ & 0.072 \\ 
\textbf{Latin} & $ 14.9 \pm 8.8 $ & $ 0.169 \pm 0.113 $ &       $ 0.083 \pm 0.044 $ & 0.028 \\  
\textbf{Persian} & $ 14.5 \pm 10.9 $ & $ 0.148 \pm 0.103 $ &    $ 0.088 \pm 0.048 $ & 0.026 \\
\textbf{Portuguese} & $ 20.9 \pm 12.9 $ & $ 0.122 \pm 0.091 $ & $ 0.065 \pm 0.041 $ & 0.019 \\ 
\textbf{Romanian} & $ 10.7 \pm 6.2 $ & $ 0.181 \pm 0.124 $ &    $ 0.102 \pm 0.048 $ & 0.056 \\ 
\textbf{Russian} & $ 15.4 \pm 8.6 $ & $ 0.140 \pm 0.103 $ &     $ 0.080 \pm 0.041 $ & 0.025 \\ 
\textbf{Slovak} & $ 14.3 \pm 9.0 $ & $ 0.173 \pm 0.117 $ &      $ 0.087 \pm 0.045 $ & 0.044 \\ 
\textbf{Slovenian} & $ 16.1 \pm 10.1 $ & $ 0.200 \pm 0.126 $ &  $ 0.082 \pm 0.044 $ & 0.052 \\ 
\textbf{Spanish} & $ 27.1 \pm 13.8 $ & $ 0.111 \pm 0.085 $ &    $ 0.050 \pm 0.033 $ & 0.012 \\
\textbf{Swedish} & $ 16.2 \pm 8.9 $ & $ 0.145 \pm 0.090 $ &     $ 0.076 \pm 0.039 $ & 0.013 \\ 
\textbf{Tamil} & $ 14.6 \pm 7.4 $ & $ 0.159 \pm 0.103 $ &       $ 0.084 \pm 0.037 $ & 0.008 \\ 
\textbf{Telugu} & $ 5.4 \pm 1.6 $ & $ 0.161 \pm 0.166 $ &       $ 0.100 \pm 0.075 $ & 0.305 \\ 
\textbf{Turkish} & $ 11.8 \pm 7.9 $ & $ 0.177 \pm 0.130 $ &     $ 0.097 \pm 0.052 $ & 0.078 \\ \hline
\textbf{Macro avg} & $ 16.0 \pm 9.0 $ & $ 0.157 \pm 0.110 $ &   $ 0.081 \pm 0.043 $ & 0.049 \\ \hline 
\end{tabular}
\end{center}
\end{scriptsize}
\end{table}

\begin{table}
\caption{\label{hubiness_table_prague} A summary of the hubiness in treebanks from different languages (Prague annotation). The format and data shown are as in Table \ref{hubiness_table_stanford}. }
\begin{scriptsize}
\begin{center}
\begin{tabular}{lrrrr}
\hline
Language & $n$ & $h$ & $E_{URLT}[h | \neg \mbox{star}]$ & $p(star)$ \\ \hline
\textbf{Arabic} & $ 26.4 \pm 21.0 $ & $ 0.050 \pm 0.055 $ &     $ 0.064 \pm 0.045 $ & 0.002 \\
\textbf{Basque} & $ 11.7 \pm 5.2 $ & $ 0.163 \pm 0.115 $ &      $ 0.095 \pm 0.041 $ & 0.043 \\
\textbf{Bengali} & $ 7.4 \pm 2.8 $ & $ 0.232 \pm 0.160 $ &      $ 0.121 \pm 0.049 $ & 0.168 \\
\textbf{Bulgarian} & $ 13.4 \pm 7.6 $ & $ 0.104 \pm 0.101 $ &   $ 0.089 \pm 0.045 $ & 0.021 \\
\textbf{Catalan} & $ 26.9 \pm 13.9 $ & $ 0.073 \pm 0.067 $ &    $ 0.049 \pm 0.031 $ & 0.006 \\
\textbf{Czech} & $ 16.2 \pm 8.7 $ & $ 0.092 \pm 0.081 $ &       $ 0.075 \pm 0.040 $ & 0.013 \\
\textbf{Danish} & $ 17.1 \pm 9.8 $ & $ 0.096 \pm 0.089 $ &      $ 0.074 \pm 0.041 $ & 0.018 \\
\textbf{Dutch} & $ 13.5 \pm 8.3 $ & $ 0.110 \pm 0.103 $ &       $ 0.090 \pm 0.047 $ & 0.019 \\
\textbf{English} & $ 21.7 \pm 10.0 $ & $ 0.086 \pm 0.069 $ &    $ 0.058 \pm 0.032 $ & 0.007 \\
\textbf{Estonian} & $ 7.2 \pm 4.0 $ & $ 0.211 \pm 0.173 $ &     $ 0.111 \pm 0.062 $ & 0.236 \\
\textbf{Finnish} & $ 12.1 \pm 5.6 $ & $ 0.133 \pm 0.104 $ &     $ 0.094 \pm 0.040 $ & 0.021 \\
\textbf{German} & $ 17.0 \pm 9.3 $ & $ 0.091 \pm 0.076 $ &      $ 0.072 \pm 0.039 $ & 0.030 \\
\textbf{Greek (Anc.)} & $ 13.3 \pm 7.9 $ & $ 0.193 \pm 0.141 $ &  $ 0.093 \pm 0.044 $ & 0.040 \\
\textbf{Greek (Mod.)} & $ 22.9 \pm 12.5 $ & $ 0.077 \pm 0.065 $ &      $ 0.057 \pm 0.035 $ & 0.006 \\
\textbf{Hindi} & $ 20.4 \pm 9.2 $ & $ 0.095 \pm 0.077 $ &       $ 0.060 \pm 0.029 $ & 0.002 \\
\textbf{Hungarian} & $ 18.7 \pm 10.4 $ & $ 0.123 \pm 0.097 $ &  $ 0.070 \pm 0.039 $ & 0.015 \\
\textbf{Italian} & $ 18.4 \pm 11.8 $ & $ 0.094 \pm 0.088 $ &    $ 0.071 \pm 0.042 $ & 0.015 \\
\textbf{Japanese} & $ 10.3 \pm 6.2 $ & $ 0.084 \pm 0.102 $ &    $ 0.100 \pm 0.052 $ & 0.037 \\
\textbf{Latin} & $ 14.9 \pm 8.8 $ & $ 0.124 \pm 0.103 $ &       $ 0.083 \pm 0.044 $ & 0.024 \\
\textbf{Persian} & $ 14.2 \pm 10.7 $ & $ 0.078 \pm 0.084 $ &    $ 0.088 \pm 0.049 $ & 0.012 \\
\textbf{Portuguese} & $ 20.8 \pm 12.9 $ & $ 0.067 \pm 0.067 $ & $ 0.065 \pm 0.041 $ & 0.009 \\
\textbf{Romanian} & $ 10.6 \pm 6.2 $ & $ 0.128 \pm 0.118 $ &    $ 0.101 \pm 0.048 $ & 0.042 \\
\textbf{Russian} & $ 15.3 \pm 8.6 $ & $ 0.092 \pm 0.086 $ &     $ 0.080 \pm 0.042 $ & 0.018 \\
\textbf{Slovak} & $ 13.9 \pm 8.7 $ & $ 0.121 \pm 0.108 $ &      $ 0.088 \pm 0.045 $ & 0.037 \\
\textbf{Slovenian} & $ 15.4 \pm 9.5 $ & $ 0.150 \pm 0.122 $ &   $ 0.084 \pm 0.044 $ & 0.041 \\
\textbf{Spanish} & $ 27.0 \pm 13.8 $ & $ 0.069 \pm 0.066 $ &    $ 0.050 \pm 0.033 $ & 0.008 \\
\textbf{Swedish} & $ 16.0 \pm 8.8 $ & $ 0.112 \pm 0.085 $ &     $ 0.077 \pm 0.039 $ & 0.011 \\
\textbf{Tamil} & $ 14.6 \pm 7.4 $ & $ 0.134 \pm 0.101 $ &       $ 0.084 \pm 0.037 $ & 0.007 \\
\textbf{Telugu} & $ 5.4 \pm 1.6 $ & $ 0.156 \pm 0.167 $ &       $ 0.103 \pm 0.075 $ & 0.342 \\
\textbf{Turkish} & $ 11.3 \pm 7.6 $ & $ 0.111 \pm 0.117 $ &     $ 0.097 \pm 0.054 $ & 0.056 \\ \hline
\textbf{Macro avg} & $ 15.8 \pm 9.0 $ & $ 0.115 \pm 0.099 $ &   $ 0.081 \pm 0.044 $ & 0.043 \\ \hline
\end{tabular}
\end{center}
\end{scriptsize}
\end{table}

\section{Discussion}

\label{discussion_section}

We have clarified the issue of the scarcity of crossing dependencies. \changed{We have provided the first evidence that the actual number of crossings is significantly small.} From the perspective of planarity, the proportion of non-planar sentences can be \changed{''high''} in certain languages (e.g., Dutch) \changed{but still significantly low}. \changed{On the other hand}, the mean number of crossings per sentence is a small number, consistently with the claim that crossings in real sentences are scarce \cite{lecerf60,hays64,melcuk88,Ferrer2006d,Park2009a,Gildea2010a} even in languages where non-planar sentences abound. However, whether a number is small or large is a matter of the scale or the units of measurement \cite{Huff1954a}. Therefore, \changed{statistical testing and} a theory of crossings (Section \ref{baselines_section}) are vital. \changed{The former shows that crossings are significantly low. The latter helps to understand why and how. }  

The low number of crossings of real sentences could be trivially explained by a high hubiness, which would immediately lead to a low value of $|Q|$, the potential number of crossings. 
Fig. \ref{hubiness_figure} indicates that this is unlikely to be the case for sufficiently large trees: the hubiness of trees tends to decrease as $n$ increases and so the relative number of crossings does (Fig. \ref{relative_crossings_figure}). The contribution of hubiness to keeping the number of crossings low decreases as $n$ increases.
 
Furthermore, the hubiness coefficient never exceeds $25\%$ and is about $10\%$ on average\changed{, although it is significantly high with respect to URLTs in the majority of treebanks}. The point: is this number large enough to expect a low number of crossings? Thanks to Eq. \ref{hubiness_vs_potential_crossings_equation}, the relative potential of crossings with respect to a linear tree turns out to be at least $75\%$, and $90\%$ on average. This strongly suggests that hubiness has a secondary role in explaining the scarcity of crossing dependencies. Indeed, we have seen above that various baselines indicate that real trees are close to linear trees. We have also seen that the gap between real trees and URLTs reduces with Prague annotations. The statistical similarity between real dependency trees and linear trees is what makes the low number of crossing dependencies to be really scarce: linear trees maximize the potential number of crossings, as we have shown above.  

The challenge for future research is to determine the true reason for the low number of crossings in sentences. A long standing hypothesis is that the low number of crossings of real sentences is a side effect of the principle of dependency length minimization, namely, the minimization of the distance between linked vertices in the linear sequence \cite{Ferrer2006d, Ferrer2014f, Ferrer2014c,Gomez2016a}. The low hubiness of real sentences suggests that hubiness may have a secondary role in reducing crossing dependencies. We hope that our quantification of the number of crossing dependencies with respect to baselines stimulates further research on the actual origin of their scarcity and the weight of different factors.  

We have observed a breakpoint in the decay of the average number of crossings across treebanks at $n = 13$ (Fig. \ref{crossings_figure}) that is also suggested by the decay of the average relative number of crossings (Fig. \ref{relative_crossings_figure}). We suspect that it could be related to increasing pressure for dependency length minimization for longer sentences. However, the real nature of the breakpoint should be investigated further.


Although the conclusion that crossings in sentences are really scarce does not depend on the annotation format, our analyses indicate that Stanford and Prague dependencies are not statistically equivalent. For instance, we have seen that real trees are closer to URLTs with respect to hubiness when Prague dependencies are considered. This is in line with recent results highlighting various other relevant quantifiable differences between annotation criteria, e.g. in their suitability for automatic parsing \cite{Rosa2015,Rehbein2017} or in the prevalence of certain patterns of crossing dependencies \cite{GomCL2016}. Thus, considering more than one annotation format is useful to analyze underlying properties of syntax, and distinguish them from properties of a specific annotation.

It is worth bearing in mind that syntactic annotation schemes are typically designed based on linguistic considerations \cite{Ide2017}, as well as technical considerations to facilitate the work of parsers and other language processing systems \cite{Schwartz2012}, independently from statistical considerations \cite{Ferrer2002f}. Our findings suggest that statistical implications should be involved when improving current annotation formats or developing new ones. Identifying the most appropriate statistical ensemble for syntactic dependency trees is an important problem that should be the subject of future research.

\section{Conclusion}

We have shown that the number of crossings of real sentences is really scarce with the help of different baselines. Although that scarcity could be easily explained by a high hubiness, the hubiness of real sentences is rather low suggesting that it has a secondary role in the low number of crossings of real sentences. Statistically, syntactic dependency trees seem to be closer to linear trees than to star trees.
Our findings provide support for the hypothesis that dependency length minimization is the main force responsible for the scarcity of crossing dependencies. 

\appendix

\section{The maximum number of crossings of a linear tree}

\label{maximum_crossings_linear_tree_appendix}


\begin{figure}
\begin{center}
\includegraphics{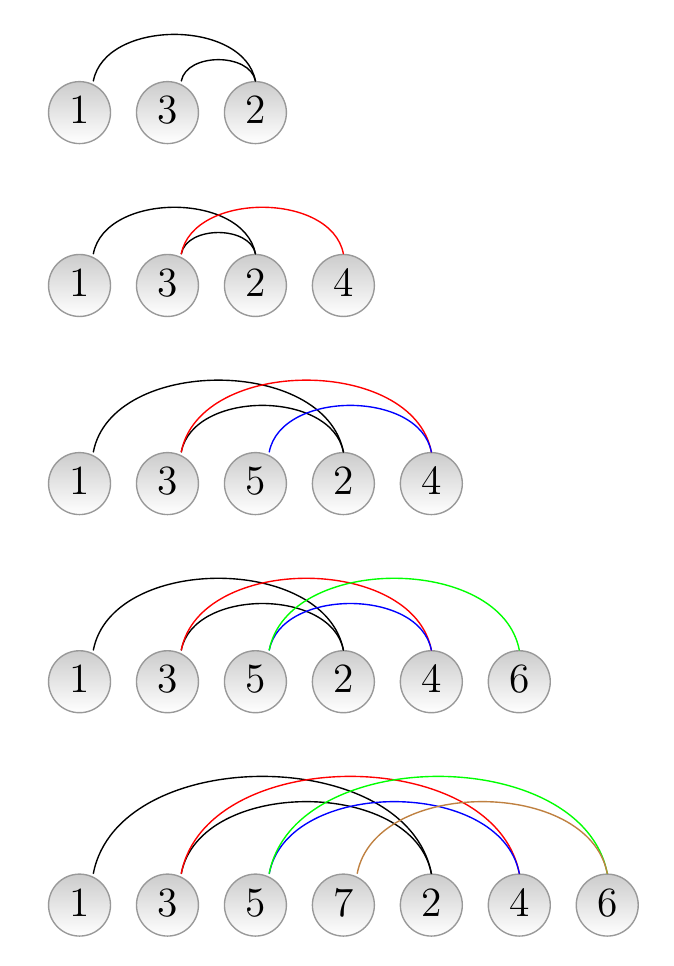} 
\end{center}
\caption{\label{extremal_arrangements_linear_tree_figure} Arrangements of linear trees that maximize the number of crossings. Top to bottom: linear trees of $3$, $4$, $5$, $6$ and $7$ nodes that have 0, 1, 3, 6 and 10 crossings, respectively. It is easy to check that $C = |Q_{linear}|$ in all cases (recall Eq. \ref{potential_number_of_crossings_of_linear_tree_equation}).}
\end{figure}


Figure \ref{extremal_arrangements_linear_tree_figure} shows arrangements with maximum number of crossings for a series of linear trees of $n$ nodes, with $3\le n\le 7$. Each tree of $n$ nodes is obtained by adding the vertex $n$ to the tree of $n-1$ nodes. In all cases, the linear ordering of the vertices consists of the odd vertex labels in increasing order, followed by the even vertex labels also in increasing order. We will show that this kind of arrangements achieves the maximum possible number of crossings for linear trees of $n$ nodes.
Formally, these orderings can be defined as the sequence of vertices 
\begin{equation}
1,3,\cdots,n+n \bmod 2-1,2,4,\cdots,n-n \bmod 2.
\end{equation}
Let $C(n)$ be the corresponding number of crossings. Notice that $C(n)=0$ for $0 \leq n\leq3$ \cite{Ferrer2013d}.

In Figure \ref{extremal_arrangements_linear_tree_figure}, we adopt the convention that the edge $3\sim 4$ is always red, $4\sim 5$ is always blue, $5\sim 6$ is always green and $6\sim 7$ is always brown for all linear trees. Thus, it is easy to check the contribution to $C(n)$ of the edge $(n-1)\sim(n)$ with respect to $C(n-1)$; when $n = 4$, the edge $3\sim 4$ adds one crossing;  when $n=5$, the edge $4\sim 5$ adds two crossings; when $n = 6$, the edge $5\sim 6$ adds three crossings and, when $n = 7$, the edge $6\sim 7$ adds four crossings. 
After this introduction now comes the proof.

We aim to show that $C(n)=|Q_{linear}|$ (Eq. \ref{potential_number_of_crossings_of_linear_tree_equation}) for $n \geq 3$. 
First, $C(3) = |Q_{linear}| = 0$, setting the base case. 
Second, we aim to show that $C(n) = \Delta(n) + C(n-1)$ with $\Delta(n) = n - 3$ for $n \geq 4$. Suppose that a tree of $n - 1$ vertices becomes a tree of $n$ vertices adding vertex $n$ and the edge $(n-1)\sim n$.
If $n$ is odd, the edge $(n-1)\sim n$ crosses any two edges formed with node $i$, namely edges $(i-1)\sim i$ and $i\sim (i+1)$, for $i$ even and $2\le i\le n-3$. This yields $\Delta(n) = n-3$. Note that $(n-1)\sim n$ cannot cross $(n-2)\sim (n-1)$ as they share vertex $n-1$.
If $n$ is even, then $(n-1)\sim n$ crosses $1\sim 2$ and any two edges formed with node $i$ such that $i$ is odd and $3\le i\le n-3$, giving again $\Delta(n) = n-3$.
Therefore, 
\begin{eqnarray}
C(n) & = & \sum_{i=4}^n \Delta (i) \nonumber \\ 
     & = & \sum_{i=1}^{n - 3} i \nonumber \\
     & = & \frac{1}{2}(n-2)(n-3)
\end{eqnarray}
and finally $C(n)=|Q_{linear}|$ (Eq. \ref{potential_number_of_crossings_of_linear_tree_equation}), as we wanted to prove. 

\section{Expectations on uniformly random labelled trees excluding star trees}

\label{expectations_without_star_trees_appendix}

There are $n$ labelled star trees: each can be constructed by choosing one of the $n$ vertices as the hub. Since there are $n^{n-2}$ labelled trees in total \cite{Cayley1889a}, the probability that a URLT is a star tree is
\begin{eqnarray}
p(\mbox{star}) & = & \frac{n}{n^{n-2}} \nonumber \\
        & = & n^{3 - n}. \label{probability_of_star_equation}        
\end{eqnarray}
We define the sum of squared degrees of a tree as \cite{Ferrer2013b}
\begin{equation}
K_2 = n \left<k^2 \right>
\end{equation}
and define $p(K_2 | \neg \mbox{star})$ as the probability that a URLT has $K_2$ as sum of squared degrees knowing that it not a star tree.
We have that 
\begin{equation}
p(K_2 | \neg \mbox{star}) = \frac{p(\neg \mbox{star} | K_2)p(K_2)}{p(\neg \mbox{star})}.
\end{equation}

\changed{We have seen above that the maximum value of $\left<k^2 \right>$ for a given $n$ is achieved by a star tree (Eq. \ref{k2_variability_range_equation}), and hence the same can be said about the maximum value of $K_2$. If we call this value $K_2^{star}$, then 
\begin{equation}
p(\neg \mbox{star} | K_2) = 
   \left\{ 
      \begin{array}{l} 
      1 \mbox{~for~} K_2 < K_2^{star} \\
      0 \mbox{~for~} K_2 = K_2^{star}.
      \end{array}
   \right.
\end{equation}
Therefore, for $K_2 < K_2^{star}$, we can apply $p(\neg \mbox{star}) = 1 - p(\mbox{star})$ and $p(\neg \mbox{star} | K_2) = 1$ to obtain}
\begin{equation}
p(K_2 | \neg \mbox{star}) = \frac{p(K_2)}{1 - p(\mbox{star})}.
\end{equation}

If star trees are excluded, the maximum hubiness is reached by a quasi-star tree, a tree that gives the second largest value of $\left<k^2\right>$, and is defined by  one vertex of degree $n-2$, one vertex of degree 2 and the remainder of vertices of degree 1 \cite{Ferrer2014f} (Fig. \ref{star_and_linear_trees_figure}).
Suppose that $K_2^{linear}$ and $K_2^{quasi-star}$ are the values of $K_2$ of a linear tree and a quasi-star tree, respectively. 
The expectation of $\left<k^2 \right>$ of a URLT knowing that it is not a star tree is
\begin{eqnarray}
E_{URLT}\left[\left<k^2 \right> | \neg \mbox{star} \right] & = & \frac{1}{n} E_{URLT}\left[ K_2 | \neg \mbox{star} \right] \nonumber \\
                                                           & = & \frac{1}{n} \sum_{K_2 = K_2^{linear}}^{K_2^{quasi-star}} p(K_2 | \neg \mbox{star}) K_2 \nonumber \\
                                                           & = & \frac{1}{n(1 - p(\mbox{star}))}\sum_{K_2 = K_2^{linear}}^{K_2^{quasi-star}} p(K_2) K_2 \nonumber \\
& = & \frac{1}{n(1 - p(\mbox{star}))} \left(E_{URLT}[K_2] - p(\mbox{star})K_2^{star} \right).
\end{eqnarray}
Knowing that 
\begin{equation}
K_2^{star} = n \left< k^2 \right>_{star} = n(n-1),
\end{equation}
\begin{eqnarray}
E_{URLT}[K_2] & = & nE_{URLT}[\left< k^2 \right>] \nonumber \\
              & = & \frac{n(n-1)(5n-6)}{n^2}
\end{eqnarray}
thanks to Eq. \ref{expected_degree_2nd_moment_equation},
and recalling Eq. \ref{probability_of_star_equation}, one obtains
\begin{equation}
E_{URLT}\left[\left<k^2 \right> \middle| \neg \mbox{star} \right] = \frac{n-1}{1 - n^{3 -n}} \left(
\frac{5n -6}{n^2} - n^{3 - n} \right).
\label{expected_degree_2nd_moment_without_stars_equation}
\end{equation}
Notice that 
\begin{equation}
E_{URLT}\left[\left<k^2 \right> \middle| \neg \mbox{star} \right] = E_{URLT}\left[\left<k^2 \right> \right]
\end{equation}
for sufficiently large $n$ (compare Eqs. \ref{expected_degree_2nd_moment_equation} and \ref{expected_degree_2nd_moment_without_stars_equation}).

Adapting Eq. \ref{expected_crossings_equation} to $\doubleexpectationwithoutstars$, one obtains
\begin{equation}
\doubleexpectationwithoutstars = \frac{n}{6} \left(n - 1 - E\left[\left<k^2\right> \middle| \neg \mbox{star}\right] \right). \label{adapted_expected_crossings_equation}
\end{equation}
Plugging Eq. \ref{expected_degree_2nd_moment_without_stars_equation} to \ref{adapted_expected_crossings_equation},
one obtains
\begin{equation}
\doubleexpectationwithoutstars = \frac{(n-1)(n-2)(n-3)}{6(n - n^{4-n})} \label{adapted_expected_number_of_crossings_of_uniformly_random_tree_equation}
\end{equation}
and also  
\begin{eqnarray}
\doubleexpectationwithoutstars & = & \frac{n-1}{3(n - n^{4 - n})}|Q|_{linear} \nonumber \\
                               & \approx & \frac{1}{3}|Q|_{linear}.
\end{eqnarray}

Adapting Eq. \ref{expected_hubiness_equation} to $E_{URLT}\left[h | \neg \mbox{star} \right]$, one obtains
\begin{equation} 
E_{URLT}\left[h \middle| \neg \mbox{star} \right] = \frac{n\left(E_{URLT}\left[\left<k^2 \right> \middle| \neg \mbox{star} \right] - \left<k^2 \right>_{linear}\right)}{(n-2)(n-3)}.
\end{equation}
Note that 
\begin{eqnarray}
E_{URLT}\left[\left<k^2 \right> \middle| \neg \mbox{star} \right] - \left<k^2 \right>_{linear} & = & 
\frac{n-1}{1 - n^{3 -n}} \left(
\frac{5n -6}{n^2} - n^{3 - n} \right) - \left(4 - \frac{6}{n}\right) \nonumber \\
 & = & \frac{(n-2)(n-3)}{n}\frac{n^4 - n^n}{n^4 - n^{n+1}}
\end{eqnarray}
and then 
\begin{equation}
E_{URLT}\left[h \middle| \neg \mbox{star} \right] = \frac{n^{n-4} - 1}{n^{n-3} - 1}.
\label{adapted_expected_hubiness_equation}
\end{equation}
It is easy to see that 
\begin{equation}
E_{URLT}\left[h \middle| \neg \mbox{star}\right] \approx E_{URLT}[h] = \frac{1}{n}
\end{equation}
for sufficiently large $n$. For numerical reasons, it is convenient to use Eq. \ref{adapted_expected_hubiness_equation} till $n = n^*$ and then replace the formula simply by $1/n$. $n^*$ can be chosen as the largest value of $n$ for which Eq. \ref{adapted_expected_hubiness_equation} does not produce numerical overflows when calculating the powers. Such a critical value increases through the decomposition   
\begin{equation}
E_{URLT}\left[h \middle| \neg \mbox{star} \right] = a(n, 1) a(n, -1),
\end{equation}
with
\begin{equation}
a(n, x) = \frac{n^\frac{n-4}{2} + x}{n^\frac{n-3}{2} + x}.
\end{equation}

All the corrected expectations that we have calculated in this section require $n \geq 4$ because there are no labelled trees with $n < 4$ such that they are not star trees.

\section*{Acknowledgements}

RFC is funded by the grants 2014SGR 890 (MACDA) from AGAUR (Generalitat de Catalunya) and also
the APCOM project (TIN2014-57226-P) from MINECO. 
CGR has received funding from the European Research Council (ERC) under the European Union's Horizon 2020 research and innovation programme (grant agreement No 714150 - FASTPARSE) and from the TELEPARES-UDC project (FFI2014-51978-C2-2-R) from MINECO.
JLE is funded by the project TASSAT3 (TIN2016-76573-C2-1-P) from MINECO (Ministerio de Econom\'{\i}a y Competitividad). 

\bibliographystyle{unsrt}
\bibliography{../../scarcity_of_crossing_dependencies/main,../../scarcity_of_crossing_dependencies/twoplanaracl,../../scarcity_of_crossing_dependencies/Ramon}

\end{document}